\documentclass[12pt]{article}

\usepackage{titlesec} 
\usepackage{amsmath}
\usepackage{graphicx}
\usepackage{multirow}
\usepackage{bm}
\usepackage{amssymb}
\usepackage{wrapfig}
\usepackage{floatflt}
\usepackage{float}
\usepackage{natbib} 
\usepackage{url} 
\usepackage{caption}
\usepackage{subcaption}
\usepackage{hyperref}
\hypersetup{
    colorlinks=true, 
    linkcolor=blue,  
    citecolor=blue, 
    urlcolor=blue    
}
\usepackage{booktabs}

\usepackage{subcaption}

\usepackage{setspace}

\newcommand{\x}{\mathbf{x}}
\newcommand{\y}{\mathbf{y}}

\newcommand{\mbr}[1]{\left\{#1\right\}}
\newcommand{\sbr}[1]{\left(#1\right)}

\newcommand{\dataset}{\mathcal{D}}

\newcommand{\given}{\middle|}

\usepackage[dvipsnames]{xcolor}
\usepackage{mathtools} 

\newcommand{\z}{\mathbf{z}}
\newcommand{\f}{\mathbf{f}}
\renewcommand{\t}{\mathbf{t}}
\usepackage{bm}
\newcommand{\bb}{\bm{b}}
\newcommand{\brho}{\bm{\rho}}
\newcommand{\bgamma}{\bm{\gamma}}

\newcommand{\e}{\bm{e}}
\newcommand{\w}{\bm{w}}
\renewcommand{\u}{\mathbf{u}}
\newcommand{\prior}{\pi}

\newcommand{\transpose}{{^\top}}

\usepackage{xspace} 
\newcommand{\ours}{\textit{MFGPCox}\xspace}
\newcommand{\deepbrch}{\textit{DeepBrch}\xspace}
\newcommand{\nnjoint}{\textit{NNJoint}\xspace}
\newcommand{\deepsurv}{\textit{DeepSurv}\xspace}
\newcommand{\coxph}{\textit{CoxPH}\xspace}

\usepackage{arxiv}

\usepackage[utf8]{inputenc} 
\usepackage[T1]{fontenc}    
\usepackage{hyperref}       
\usepackage{url}            
\usepackage{booktabs}       
\usepackage{amsfonts}       
\usepackage{nicefrac}       
\usepackage{microtype}      
\usepackage{lipsum}
\usepackage{graphicx}
\graphicspath{ {./images/} }

\date{}

\title{Bayesian Joint Model of Multi-Sensor and Failure Event Data for Multi-Mode Failure Prediction}

  \author{Sina Aghaee Dabaghan Fard
  \\
    Wm Michael Barnes ’64 Department of Industrial and Systems Engineering,\\ Texas A\&M University, College Station, Texas, 77843\\
    \And
    Minhee Kim \\
    Department of Industrial and Systems Engineering,\\ University of Florida, Gainesville, FL 32603\\
    \And
    Akash Deep \\
    School of Industrial Engineering and Management,\\
    Oklahoma State University, Stillwater, OK 74078\\
    \And
    Jaesung Lee
    \thanks{Corresponding Author: j.lee@tamu.edu, https://orcid.org/0000-0001-7533-1865 (ORCID)}\hspace{.2cm}
    \\
    Wm Michael Barnes ’64 Department of Industrial and Systems Engineering,\\  Texas A\&M University, College Station, Texas, 77843\\} 

\begin{document}
\maketitle

\vspace{-2em}

\noindent\textit{This is an original manuscript of an article published online by Taylor \& Francis in \textit{Technometrics} on May 13, 2026, available at: \url{https://doi.org/10.1080/00401706.2026.2653564}.}

\vspace{2em}

\begin{abstract}
Modern industrial systems are often subject to multiple failure modes, and their conditions are monitored by multiple sensors, generating multiple time-series signals. Additionally, time-to-failure data are commonly available. Accurately predicting a system's remaining useful life (RUL) requires effectively leveraging multi-sensor time-series data alongside multi-mode failure event data. In most existing models, failure modes and RUL prediction are performed independently, ignoring the inherent relationship between these two tasks. Some models integrate multiple failure modes and event prediction using black-box machine learning approaches, which lack statistical rigor and cannot characterize the inherent uncertainty in the model and data. This paper introduces a unified approach to jointly model the multi-sensor time-series data and failure time concerning multiple failure modes. 
This proposed model integrates a Cox proportional hazards model, a Convolved Multi-output Gaussian Process, and multinomial failure mode distributions in a hierarchical Bayesian framework with corresponding priors, enabling accurate prediction with robust uncertainty quantification. Posterior distributions are effectively obtained by Variational Bayes, and prediction is performed with Monte Carlo sampling.
The advantages of the proposed model are validated through extensive numerical and case studies with a jet-engine dataset.
\end{abstract}

\section{Introduction}
\label{sec:intro}

Failure prognostics and diagnostics are essential for reliability management in modern industrial systems, which degrade over time and often fail due to various causes. Prognostics is concerned with predicting failure time, whereas diagnostics focuses on analyzing the root cause, also called the failure mode. Remaining Useful Life (RUL), defined as the time until the failure event \citep{kim2017prognostics}, is often estimated. The probability distribution of RUL is typically characterized by the survival function, the probability that the physical system unit remains operational beyond a specified future time \citep{cox1984analysis}, thereby providing comprehensive information regarding the uncertainty associated with the failure event.
Different failure causes generally involve distinct degradation patterns, leading to corresponding variations in failure times. Therefore, comprehensive modeling and predictions of both failure time and the underlying root causes (failure modes) facilitate not only accurate failure time estimation but also effective proactive maintenance planning.

Prognostic models often rely on three key data sources: failure event times, failure modes, and condition monitoring (CM) sensor signals. Failure times are commonly available in industrial logs, generally with failure modes. Sensor technological advancements have made CM sensor data increasingly available. CM signals track physical system units' health, providing real-time insights into their degradation and enabling online updating of the failure prediction. Throughout this paper, we refer to each monitored physical system unit of concern as a \emph{unit}. Utilizing both event and sensor data ensures accurate predictions of a unit's failure time and mode, and continuous updates as sensor data is collected.

Prognostic models can be grouped into three categories: event time-based, CM signal-based, and joint models. Event time-based models, including survival models, such as variants of the Cox Proportional Hazards (PH) model, use event times with time-invariant covariates (e.g., equipment manufacturer \citep{zhou2014remaining}). While most existing event time-based models only consider a single failure mode \citep{cox1972regression, katzman2018deepsurv}, there are some models that consider multiple failure modes, such as competing risks approaches \citep{lau2011parametric,nagpal2021deep}. Nevertheless, a critical limitation of the event time-based models is that they cannot capture the most recent real-time information of the systems' degradation, which limits their real-time updating and predictive accuracy. On the other hand, CM signals-based models solely rely on time series sensor signals and assume that failures are when a degradation signal surpasses a predefined threshold (also called soft failure); however, such an oversimplified definition of failure is often not consistent with the real failure events (hard failure) \citep{gebraeel2005residual,liu2013data,liu2014integration}. 

Joint models integrate the CM signal-based models into the event time-based models. Most works first predict CM signals and plug the point prediction into the survival model for failure prediction. However, point prediction tends to introduce biases, and the prediction uncertainties are overlooked in such separate two-step approaches \citep{zhou2014remaining, yue2021joint}. 
A joint model called NN-Joint predicts the CM signals by a linear mixed-effects model and uses its predictions as an input of a neural network-based Cox PH to predict the survival function \citep{wen2023neural}; CM signals are assumed to follow the quadratic relationship with respect to times; however, such a parametric assumption is often too rigid in practice as sensor data usually do not follow specific parametric forms. Nonparametric approaches, such as Functional Principal Component Analysis (FPCA) with a Bayesian neural network \citep{huh2024integrated} or Convolved Multiple Output Gaussian Processes (CMGP) coupled with Cox PH \citep{yue2021joint} offer greater flexibility. However, these and most of the aforementioned methods consider only a single failure mode and are unable to characterize different failure modes.

Multiple failure modes have been considered in joint models. They can also be categorized into two groups: two-step methods and unified models, depending on whether failure mode classification and RUL prediction are performed in separate steps or simultaneously. In two-step approaches, first, failure modes are classified based on observed data, and for the identified failure mode, RUL prediction is made \citep{chehade2018data}. Two-step models overlook classification uncertainty, leading to biased and overconfident RUL estimates \citep{zhang2021survey}. Unified models predict failure time and mode simultaneously within a single model. Existing unified models are based on neural networks due to the high problem complexity. In \cite{10032096}, a linear mixed-effects model is integrated with a deep neural network for failure mode classification and RUL prediction, where the linear mixed-effects model is not flexible. Another unified model, Deep Branched Network (DeepBrch), employs Long Short-Term Memory (LSTM) and a fully connected neural network architecture for both failure time and mode prediction \citep{9845469}. 
However, both of these models only provide a single point estimate of RUL and are not able to predict the survival probability at a given future point. Additionally, such black-box neural network-based models lack interpretability, require large data sizes, lack statistical rigor, and cannot quantify uncertainty.

In this paper, we propose the Multi-Failure Gaussian Process Cox Model (MFGPCox) to address the aforementioned challenges. The primary contributions of this paper are as follows:

\begin{itemize}
\item The proposed model simultaneously predicts failure modes and failure times by leveraging failure event time and mode data, and CM signals from multiple sensors within a unified Bayesian hierarchical framework. Within MFGPCox, CMGP captures both the global trends and unit-to-unit variations of CM signals, while Cox PH accounts for the shared characteristics through the baseline hazard and the individual variations by the covariates in the failure event data. By explicitly modeling each data source while sharing information through the hierarchy, MFGPCox achieves high interpretability and enables accurate and personalized predictions of failure modes and RUL, which are updated over time. For efficient training, we use variational inference, where an evidence lower bound is derived and maximized. 
\item The proposed model precisely characterizes model uncertainty and quantifies the uncertainty in prediction in a statistically rigorous manner. 
The CMGP is inherently Bayesian, and to better quantify the uncertainty in the proposed framework, the Cox PH model is formulated using the Bayesian approach by assigning priors to its parameters. 
In survival prediction, we use Monte Carlo sampling to fully characterize the uncertainties across the time series and failure time and mode predictions. The full Bayesian nature of our proposed model ensures robust performance even with limited data. 
\item To demonstrate the benefits of the proposed model, extensive numerical and case studies have been performed. The proposed model outperforms state-of-the-art benchmark methods in three key aspects: failure mode prediction, RUL (survival probability) prediction, and uncertainty quantification.
\end{itemize}

The remainder of this paper is structured as follows. \hyperref[sec:background]{Section \ref{sec:background}} provides the necessary background on the core components of the proposed framework. The proposed model is presented in \hyperref[sec:model]{Section \ref{sec:model}}. 
Inference and prediction procedures are described in \hyperref[sec:Inference]{Section 4} and \hyperref[sec:pred]{Section \ref{sec:pred}}, respectively.
\hyperref[sec:Validation]{Section \ref{sec:Validation}} presents the numerical and case study, and \hyperref[sec:conc]{Section \ref{sec:conc}} concludes with discussion on future directions. 

\section{Background}
\label{sec:background}

Our proposed model incorporates CMGP in the Cox PH model to effectively characterize both time-to-failure and CM signal data; in this section, we review the employed Cox PH and the CMGP models.

\subsection{Cox Proportional Hazard Model}
\label{subsec:survival analysis}

Cox PH model describes the failure event time data and predicts the failure probability at given times.
The historical dataset includes $I$ units of physical systems that degrade over time. Each unit $i$ fails at time $T_i$ and its data may be censored at time $C_i$. The event time is denoted as $V_i=\min(T_i,C_i)$, where the event type is denoted with $\delta_i$ ($\delta_i = 1$ if failure or $\delta_i = 0$ if censored).
The survival probability of an $i$th unit at given time $t$, also known as survival function $S_i(t)= P(T_i \geq t)$, is commonly characterized by the hazard function $h_i(t)$, the instantaneous rate of failure occurrence \citep{cox1984analysis}, formally defined as $ h_i(t) = \lim_{\Delta \to 0} \frac{1}{\Delta} \mathbb{P}(t < T_i \le t + \Delta \mid T_i \ge t)$.
At time $t^*$, the probability that the unit survives additional \( \Delta t \) time is,
\begin{equation}
S_i(t^* + \Delta t \mid t^*) = \exp\left( -\int_{t^*}^{t^* + \Delta t} h_{i}(l) dl \right).
\label{eq:survival probability}
\end{equation}
A primary goal in the survival model is to estimate the hazard function based on the observed data.
Cox PH \citep{cox1972regression}, one of the widely used models, conventionally considers only the time-invariant covariates $\x_i$. 
Later extensions allow for the incorporation of time-variant covariates $\f_i(t)$, simply using the latest observations \citep{cox1984analysis}. 
The hazard function of Cox PH is defined as \( h_i(t) = h_0(t) \exp \{ \boldsymbol{\gamma}^\top \mathbf{x}_i + \bm{\beta}\transpose \f_i(t) \} \), where \( h_0(t) \) is the baseline hazard function, representing the shared hazard across all individual units and $\bm{\gamma}$ and $\bm{\beta}$ are the coefficients to the covariates $\x_i$ and $\f_i$ of unit $i$.
The parameters of the Cox PH are then estimated by maximizing the likelihood function.

\subsection{Sparse Convolved Multiple Output Gaussian Processes}
\label{subsection:CMGP_SP}
Multiple CM signals obtained from the same sensors often share common patterns, while individual signal exhibits individual characteristics. 
Gaussian Processes (GPs)  assume that the latent functional outcomes $\f=f(\t)$ at $\t$ ($\t,\f \in  \mathbb{R}^{N_i}$ where $N_i$ denotes the number of observations for unit $i$) follow the multivariate normal distribution \citep{rasmussen2006gaussian} and predict $\f$ once noisy observations $\y$ are obtained, by the posterior distribution $P(\f|\y)$, allowing full uncertainty quantification. 
However, the conventional GP generally assumes a single underlying function, which is limited in characterizing individual variations in CM signals.

In contrast, CMGP captures not only commonalities but also the random variations across signals \citep{alvarez2008sparse}. A function $f_i$ for $i$th unit is defined as 
\begin{equation}
\label{eq:convolution}
f_{i}(t) = \sum_{m=1}^{M} \int_{\mathbb{R}} G_{im}(t - r) \, u_{m}(r) \, dr,
\end{equation}
where $u_m(t)$ ($m=1,\ldots,M$) are the latent processes shared across units ($i=1,\ldots,I$) and modeled by the GP, and $G_{im}(\cdot)$ is a unit-specific smoothing kernel.
Many practical signals are successfully expressed by a single latent process (\( M = 1 \)), and it has yielded good performance~\citep{alvarez2011computationally, yue2021joint, chung2022weakly}. 
Multiple latent processes ($M>1$) can also be used when the target functions manifest a mixture of multiple patterns. For notational simplicity, we use $M=1$ and drop $m$ from the following equations.

\begin{wrapfigure}{r}{0.49\textwidth} 
    \vspace{-2em}
    \centering
    \includegraphics[width=.48\textwidth]{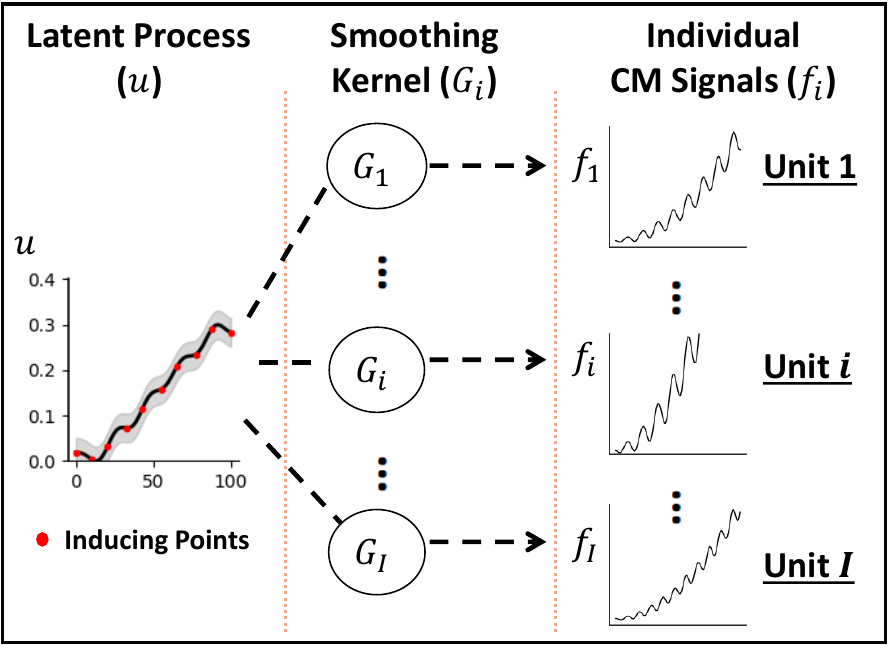}
	\caption{Convolved Multi-output GP}\label{fig:CMGP}
\end{wrapfigure}

Figure \ref{fig:CMGP} illustrates the CMGP. Individual functions ${f}_{i}$ are defined by the convolution of the latent process $ u$ and individual kernels $G_i$; the latent process $u$ captures underlying common key patterns exhibited across individual signals, while the kernel $G_i$ captures individual variations of the specific $i$th signals. By the convolution operation in \eqref{eq:convolution}, CMGP successfully characterizes both commonality and variations across units.

The Radial Basis Function (RBF) kernel is often used for the latent processes
$ K_{u_{j},u_{j}}(t, t')\\ = \exp\left(-{(t - t')^2}/{(2\lambda^2)}\right)
$ and the smoothing kernel $G_{i}(t) = \eta_{i}/{\sqrt{2\pi (\xi_{i})^2}} \exp\left(-{t^2}/{(2\xi_{i}^2)}\right)$ \citep{alvarez2008sparse}
where \(\lambda\), \(\eta_{i}\), and \(\xi_{i}\) are the length scale, scaling coefficient, and smoothing parameter, respectively. These are hyperparameters to be optimized, often by maximizing the log marginal likelihood \citep{rasmussen2006gaussian}.

Like conventional GP, one of the critical limitations of CMGP is the computational complexity, mainly induced by the inverse operation of the covariance matrix. When the total number of data points is denoted as $N=\sum_i N_i$, the computational complexity in training CMGP is \( \mathcal{O}({N}^3) \) \citep{rasmussen2006gaussian}.
The complexity can be reduced by the sparse approximation, which introduces inducing variables \(\mathbf{u} = u(\w)\), where the latent process is characterized at fixed $Q$-dimensional pseudo-inputs $\w$ and $Q<<N_i$ \citep{snelson2005sparse}. 
This approximation significantly reduces the computational complexity to \( \mathcal{O}(Q^2 N) \) while maintaining its efficacy \citep{alvarez2011computationally}.

\section{Proposed Model: MFGPCox}
\label{sec:model}

CM signals from $J$ sensors are sequentially collected over time from each physical system unit $i$ (out of $I$ units) until it fails at $T_i$ under failure mode $k$ (out of $K$ possible modes), or censored at $C_i$. The objective of this work is to predict the failure time $T_i$ and failure mode $k$ by using both the time-to-failure and CM signal data.
For each unit $i$, three sources of data are observed. The time-to-failure data consists of the event time $V_i=\min(T_i,C_i)$, event type $\delta_i$ (indicating failure or censoring), and covariates $\x_i$. The CM signal data, collected with noise from $j$th sensor at time points $\t_{ij}=(t_{ij1},\ldots,t_{ijN_i})\transpose$  is denoted as $\y_{ij}=(y_{ij1},\ldots,y_{ijN_i})\transpose$, where $N_i$ is the number of CM observations. 
The experienced failure mode is denoted as $\z_i =(z_{i1},\ldots, z_{iK})$, where $z_{ik}=1$ if the unit $i$ experiences $k$th failure mode and $z_{ik'}=0$ for all $k'\not=k$. 
Our proposed model characterizes these three heterogeneous data sources within a unified hierarchical Bayesian framework. 
The full proposed model is presented grouped by the data source:
\begin{align}
\shortintertext{\textbf{Failure Time:}}
h_i\sbr{t\given\x_i, \mbr{\f_{ij}}_j; b_k,\rho_k}&= h_0(t;b_k,\rho_k) \exp\left\{ \bgamma_k^{\top} \mathbf{x}_i + \sum_{j=1}^{J} \beta_{kj} f_{ij}(t) \right\} \quad \text{if }\z_{i}=\bm{e}_k,
\label{eq:hazard_function}
\\
b_k &\sim 
\mathcal{N}\left({\mu}_{b_k}, \sigma_{b_k}^2 \right), \text{ and }
  \rho_k \sim \Gamma \sbr{ {\alpha}_{\rho_k}, \beta_{\rho_k} } \quad 
  \label{eq:cox_priors}
\shortintertext{\textbf{CM Sensor Signals:}}
\y_{ij}  &= 
\f_{ij} + \bm{\epsilon}_{ij}, 
\label{eq:MCGP_likelihood} 
\\
\f_{ij} \mid \z_{i}  &\sim 
\mathcal{CMGP} \sbr{ \bm{0},K_{{ij},{i'j}}(t,t')}, \label{eq:f_prior}
\shortintertext{\textbf{Failure Modes:}}
\z_i | \bm{\Pi} &\sim \text{Multinomial}(\bm{\Pi}) 
\label{eq:z} 
\\
\bm{\Pi} &\sim \text{Dir}(\bm{\alpha}) \label{eq:pi}
\end{align}
defined for every unit $i$, sensor $j$, and failure mode $k$ where $\bm{e}_k$ is a unit vector where $k$th value is 1 and other values are 0.
Equations~\eqref{eq:hazard_function}-\eqref{eq:pi} define the proposed model, which jointly captures failure times, CM signals, and failure modes. 
The hazard function in \eqref{eq:hazard_function} enables to obtain the survival probability by \eqref{eq:survival probability}, based on individual units' characteristics described by the time-invariant covariates \( \mathbf{x}_i \) and the latent denoised CM signals \( \f_{ij} \). Notice that the input $\f_{ij}$ in the hazard function \eqref{eq:hazard_function} is a random vector, ensuring that the uncertainties in CM signals carry forward to the survival model to fully account for the uncertainty in the failure time prediction. The model employs failure-mode specific coefficients ($\bm{\gamma}_k$ and $\bm{\beta}_k$) to account for the varying associations between the hazard and covariates across failure modes, as different failure mechanisms typically exhibit distinct hazard-input relationships. 
Exponential function is employed for baseline hazard function, which has a desirable monotonically non-decreasing property \citep{yue2021joint}: $h_0(t) = \exp(b_k + \rho_k t)$ for $\z_i=\bm{e}_k$.
To fully characterize the model uncertainty in the Bayesian framework, priors are assigned to the baseline hazard parameters \( b_k \) and \( \rho_k \) as defined in \eqref{eq:cox_priors}. To ensure a non-decreasing baseline hazard, \( \rho_k \) needs to be non-negative and is assigned a Gamma prior with shape and and scale parameters, \( \alpha_{\rho_k} \) and \( \beta_{\rho_k} \), respectively, while \( b_k \) is unconstrained and assigned a Gaussian prior with mean \( \mu_{b_k} \) and variance \( \sigma^2_{b_k} \).

The observed CM signal $\y_{ij}$ is modeled using a latent function $f$ with additive measurement error $\epsilon$ that follows the zero-mean normal distribution with variance $\sigma^2_\epsilon$ in \eqref{eq:MCGP_likelihood}.
The denoised latent signal $\f_{ij}=f(\t_{ij})$ is modeled by CMGP in \eqref{eq:f_prior}. \(K_{ij,i'j}(t, t')\) represents the kernel function between signals \(\mathbf{f}_{ij}(t)\) and \(\mathbf{f}_{i'j}(t')\) from sensor \(j\), and its derivation is included in Section S1 of the supplementary material. CMGP characterizes individual variations across units while the common characteristics of the CM signals are captured by the latent process as described in Section \ref{subsection:CMGP_SP}. 
For each failure mode $k$ and sensor $j$, dedicated latent processes are assigned.
Failure modes are modeled by a latent variable \( \mathbf{z}_i \), which follows a multinomial distribution parameterized by \( \bm{\Pi} = (\Pi_1, \ldots, \Pi_K) \), representing the failure mode probabilities as shown in \eqref{eq:z}, and a Dirichlet prior parameterized by \( \boldsymbol{\alpha} = (\alpha_{1}, \ldots, \alpha_{K})\) is placed on \( \bm{\Pi} \), as described in \eqref{eq:pi}. 
Please note that all three data types are represented by Bayesian models (i.e., Bayesian survival model, CMGP, and multinomial with a Dirichlet prior), which fully capture the model uncertainties in each data type, and they are successfully carried forward via conditional probabilities.

\begin{wrapfigure}{r}{0.48\textwidth} 
    \vspace{-2em}
    \centering
    \includegraphics[width=0.46\textwidth]{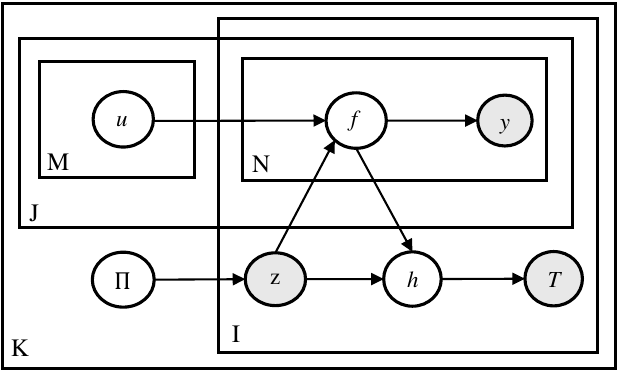}
	\caption{Simplified plate notation of the MFGPCox model. 
    Shaded nodes indicate observed values. 
    }\label{fig:plate}
\end{wrapfigure}

Figure \ref{fig:plate} clearly illustrates the relationships among key variables using plate notation.
Both the hazard function $h$ and the denoised CM signal $f$ are conditionally dependent on the experienced failure mode $z$, where $h$ governs the distribution of failure time $T$ and $f$ generates the noisy CM observation $y$.
This dependency structure facilitates inference of the failure modes a unit is experiencing based on the observations ($y$, $T$, and $z$). 
In the employed CMGP, the underlying temporal dynamics and general patterns of $f$ are captured by the latent process $u$, which is an auxiliary variable to be marginalized out.

\begin{figure}[ht!]
    \centering
    \includegraphics[width=\textwidth]{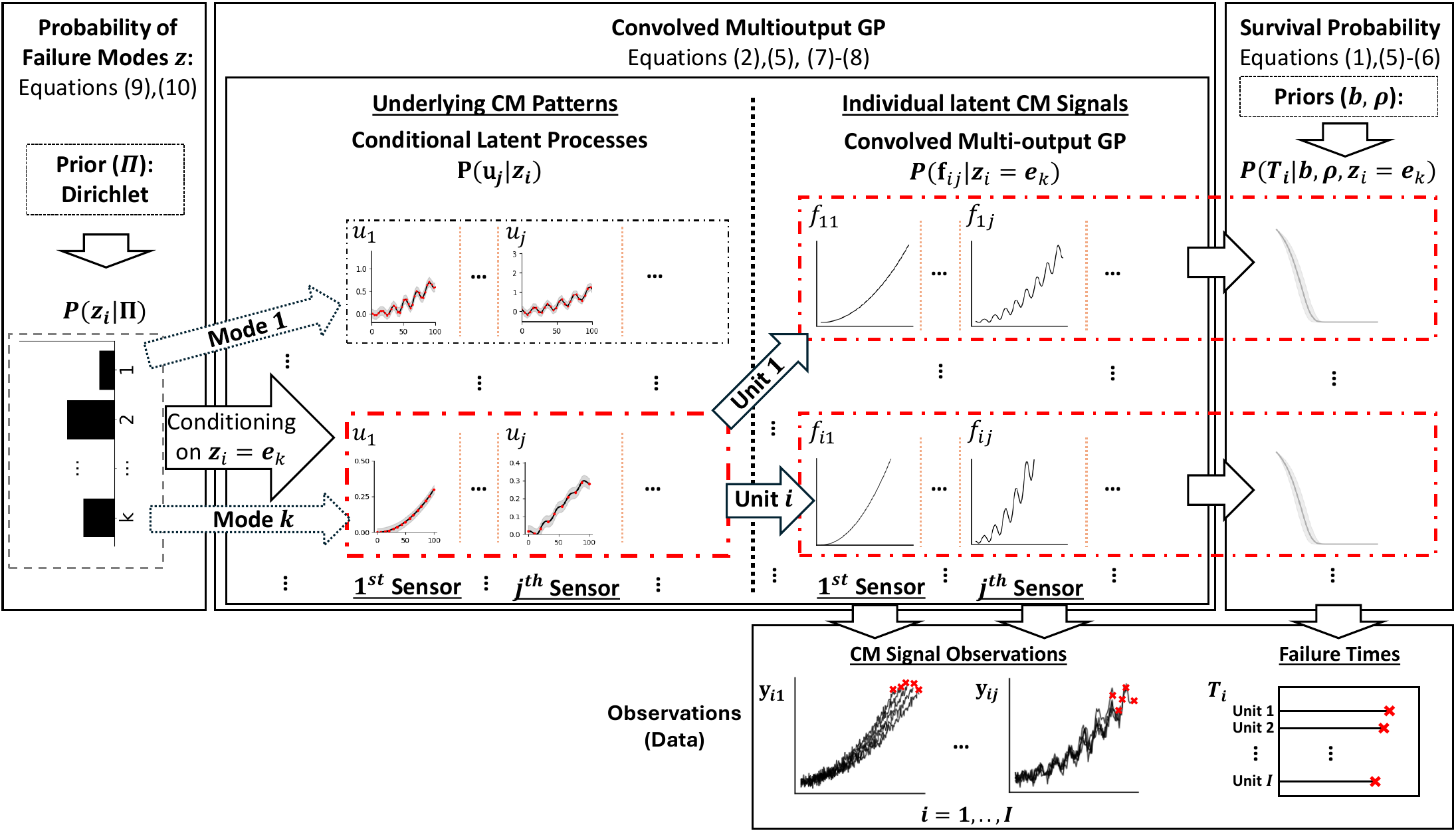}
	\caption{An overview of the MFGPCox model structure.\label{fig:model}}
\end{figure}

The overall structure of the proposed model is illustrated in Figure \ref{fig:model}. 
The failure modes are categorical probability distributions. Conditioning on both failure modes $k$ and sensors $j$, latent processes $u$ characterize the common shapes of CM signals. CMGP accounts for unit-to-unit variations of underlying CM signals $\f_{ij}$ successfully, which are then indirectly measured with noise as $\y_{ij}$. With the recovered individual latent CM signals $\f_{ij}$ and individual covariates $\x_{i}$, the Cox model predicts conditional survival probability on failure modes, which are observed with failure times.

The proposed model (MFGPCox) comprehensively models three types of data sources within a unified Bayesian hierarchical framework, while taking advantage of the benefits of Cox PH models and CMGP in accounting for individual variations and common characteristics. 
The hierarchical Bayesian structure of the model allows each component to model its respective data type effectively, resulting in a highly flexible and interpretable framework.

\section{Inference}
\label{sec:Inference}

In Bayesian models, the parameters of the likelihood functions assigned with prior distributions, denoted as $\theta=\mbr{\f, \boldsymbol{\Pi}, \bb, \brho}=\mbr{\mbr{\f_{ij}}_{i,j}, \boldsymbol{\Pi}, \mbr{\sbr{b_k, \rho_k}}_k}$, are referred to as latent variables in this paper.
The latent variables are inferred as the posterior distribution given the dataset $\dataset=\mbr{\mathbf{V}, \bm{\delta}, \x, \t, \y, \z}$, consisting of three different data sources, time-to-failure ($\sbr{\mathbf{V}, \bm{\delta}, \x}=\mbr{\sbr{V_i, \delta_i, \x_i}}_i$), CM observations ($\sbr{\t, \y}= \mbr{\sbr{t_{ij}, y_{ij}}}_{i,j}$), 
and failure modes $\z=\mbr{\z_i}_i$.
Additionally, $\u=\mbr{\u_j}_j$ is an auxiliary variable to be marginalized out. 
When prior distributions involve hyperparameters, they are often optimized. Our model has the CMGP hyperparameters, denoted as $\Psi=\mbr{\mbr{\lambda_{kj}}, \mbr{(\eta_{kij}, \xi_{kij})}_i}_{k,j}$.
We first optimize the hyperparameters $\Psi$ and then obtain the posterior distribution of $\theta$.
The CMGP hyperparameters $\Psi$ are optimized by maximizing the log marginal likelihood $\log p(\mathcal{\bm D})$ \citep{rasmussen2006gaussian}, which is 
\begin{equation}
\begin{aligned}
\log \int_{\f,\u,\bm \Pi,\bb, \brho} 
p(\mathbf{V}, \y, \z \mid \bm{\delta}, \f, \x, \boldsymbol{\Pi}, \bb, \brho) 
\prior(\f \mid  \u, \mathbf{z}) \prior(\u|\z)
\prior(\boldsymbol{\Pi}) \prior(\bb) \prior(\brho)
d\f d\u d\boldsymbol{\Pi} d\bb d\brho
\end{aligned}
\label{eq:marginal_log_joint_likelihood}
\end{equation}
where $p(\mathbf{V}, \y, \z \mid \bm{\delta}, \f, \x, \boldsymbol{\Pi}, \bb, \brho)$ is the joint likelihood of observed data (failure time and mode, and CM signals), and the rest terms $\prior(\cdot)$ are priors over its parameters. 
$p(\mathbf{V}, \y, \z \mid \bm{\delta}, \f, \x, \boldsymbol{\Pi}, \bb, \brho)$ is multiplication of three likelihood terms, $p(\mathbf{\bm{V}} \mid \bm{\delta}, \mathbf{f},\mathbf{x}, \mathbf{z}, \mathbf{b}, \bm{\rho}) p(\mathbf{\bm{y}} \mid  \mathbf{f}) p(\z|\boldsymbol{\Pi})$.

However, Equation~\eqref{eq:marginal_log_joint_likelihood} is intractable due to the integrals, and a closed-form is not available over $\Psi$. Moreover, due to the dependencies between failure time and $J$ different sensor signals, $\Psi$ needs to be optimized with the whole $(I\times J)$ CM signals under each failure mode.
To address this issue, when optimizing $\Psi$, we decouple the weak dependencies between $V_i$ and $\f_{ij}$, while focusing on the strong relation between $\y_{ij}$ and $\f_{ij}$. 
Such a modular inference approach is practical when joint inference is analytically or computationally intractable and one sub-model is significantly less informative than the other \citep{bayarri2009modularization}. Because 
\(\z\) are observed for historical units, their likelihood is independent of the CMGP hyperparameters and does not affect the optimization; as a result, we optimize \(\Psi\) by maximizing the marginal log-likelihood of the sparse CMGPs.
\begin{equation}
\begin{aligned}
\log p(\mathbf{y}\mid \Psi) 
= \sum_{k=1}^{K}\sum_{j=1}^{J}\log 
\int_{\mathbf{u}_{j}}&
\Bigg\{
\prod_{i=1}^{I}
\int_{\mathbf{f}_{ij}}
p(\mathbf{y}_{ij} \mid \mathbf{f}_{ij}, \Psi)\,
p(\mathbf{f}_{ij} \mid \mathbf{u}_{j},\z_i=\e_k, \Psi)\,
d\mathbf{f}_{ij}
\Bigg\} \times p(\mathbf{u}_{j}\mid \z=\e_k, \Psi)\, d\mathbf{u}_{j}.
\label{eq:CMGP_marginal_log_likelihood}
\end{aligned}
\end{equation}
where the integral in the above expression can be computed analytically, as detailed in Section~S2 of the supplementary materials.
This modular inference achieves both efficacy and efficiency. First, it allows us to use the exact inference in a closed form, enabling precise uncertainty representation. 
Second, optimizations can be achieved efficiently in parallel by optimizing the hyperparameters of whole units by each sensor and failure mode.

In theory, the posterior distributions of whole latent variables $\theta$ are obtained 
by Bayes’ rule: \(p(\theta \mid \mathcal{\bm D})= 
 {p(\mathcal{D} \mid \theta) 
\prior(\theta)}/{ 
p(\mathcal{\bm D}) }\), in which the marginal likelihood \( p(\mathcal{\bm D}) \) is not available in closed form, making the exact posterior intractable.
One common way to infer the posterior distribution is using sampling, such as Markov Chain Monte Carlo (MCMC), but they are often computationally expensive and slow in convergence \citep{gelman2013bayesian}. 
Variational Inference (VI) is often more scalable in approximating the posterior distribution \citep{blei2017variational}. VI approximates the complex and intractable posterior distribution ($p(\theta|\dataset)$) with a simpler and parameterized variational distribution ($q(\theta)$) by minimizing its discrepancies from the true posterior. VI minimizes Kullback-Leibler (KL) divergence between the variational distribution \( q \) and the true posterior distribution, defined as:
\[
\text{KL} \big(q(\mathbf{f,u}, \bm{\Pi}, \mathbf{b}, \bm{\rho}) \,\big\|\, p(\mathbf{f, u}, \, \bm{\Pi}, \mathbf{b}, \bm{\rho} \mid \mathbf{V}, \y, \z) \big)
= \mathbb{E}_{q} \left[ \log \frac{q(\mathbf{f,u}, \bm{\Pi}, \mathbf{b}, \bm{\rho})}{p(\mathbf{f,u}, \bm{\Pi}, \mathbf{b}, \bm{\rho} \mid \mathbf{V}, \y, \z)} \right],
\]

Mean-Field VI (MFVI) framework is employed, where the variational distributions over latent variables $\bm{\Pi},\bb, \brho $ are assumed to be independent \citep{blei2017variational}. 
Variational distributions \( q(\boldsymbol{\Pi};\tilde{\bm \alpha}) \), \( q(\mathbf{b}; \mbr{(\tilde{\mu}_{b_k}, \tilde{\sigma}^2_{b_k})}_k) \), and \( q(\boldsymbol{\rho};\{(\tilde{\alpha}_{\rho_k}, \tilde{\beta}_{\rho_k})\}_k) \) are assumed to belong to the same families as their corresponding priors. 
For $\f$, instead of using a variational distribution, we use the exact posterior of $\f$ after marginalizing $\u$ out, given the CM observations $\y$, obtained in closed form \citep{alvarez2011computationally}: 
\(
p(\mathbf{f} \mid \mathbf{y}, \z) = \int p(\mathbf{f} \mid \mathbf{u}, \z) \, p(\mathbf{u} \mid \mathbf{y},\z) \, d\mathbf{u}
\), which provides a tighter lower bound.
Using the exact inference for CMGP ensures precise uncertainty quantification in its posterior predictive distribution \citep{bui2017unifying}. 
The exact form of this predictive distribution and its derivation are provided in Section~S2 of the supplementary material. 

Minimizing this KL divergence is equivalent to maximizing a lower bound on the marginal log-likelihood of the data, known as the Evidence Lower Bound (ELBO) \citep{blei2017variational}.
The ELBO is obtained as:
\[
\begin{aligned}
\text{ELBO} &= \underbrace{\mathbb{E}_{p(\mathbf{f\mid y,\z=\e_k})q(\bm \Pi) q(\mathbf{b}) q(\bm{\rho})}\left[\log p(\mathbf{\bm{V}, \z} \mid  \bm{\delta}, \mathbf{f},\mathbf{x}, \boldsymbol{\Pi}, \mathbf{b}, \bm{\rho})\right]}_{\text{(1) Expected Log-Likelihood}} 
\underbrace{\text{KL}\left(q(\bm{\Pi})\,\|\,\prior(\bm{\Pi})\right)}_{\text{(2) KL Divergence: } q(\bm \Pi) \text{ vs. } \prior(\bm \Pi)}\\ &\quad - \underbrace{\text{KL}\left(q(\mathbf{b})\,\|\,\prior(\mathbf{b})\right)}_{\text{(3) KL Divergence: } q(\mathbf{b}) \text{ vs. } \prior(\mathbf{b})}  - \underbrace{\text{KL}\left(q(\bm{\rho})\,\|\,\prior(\bm{\rho})\right)}_{\text{(4) KL Divergence: } q(\bm \rho) \text{ vs. } \prior(\bm \rho)}.
\end{aligned}
\]

ELBO consists of expected log likelihood over latent variables ($\f, \bm{\Pi}, \bb, \brho$), subtracted by KL divergence between variational distributions and priors over latent variables, acting as regularizers, penalizing deviations of the variational posteriors from their corresponding priors. The exact expression of the ELBO, along with its detailed derivation and the implementation details of the inference procedure, are provided in Sections~S3 and~S5 of the supplementary material, respectively.

\section{Prediction}
\label{sec:pred}

\begin{figure}[H]
    \centering
    \includegraphics[width=\textwidth]{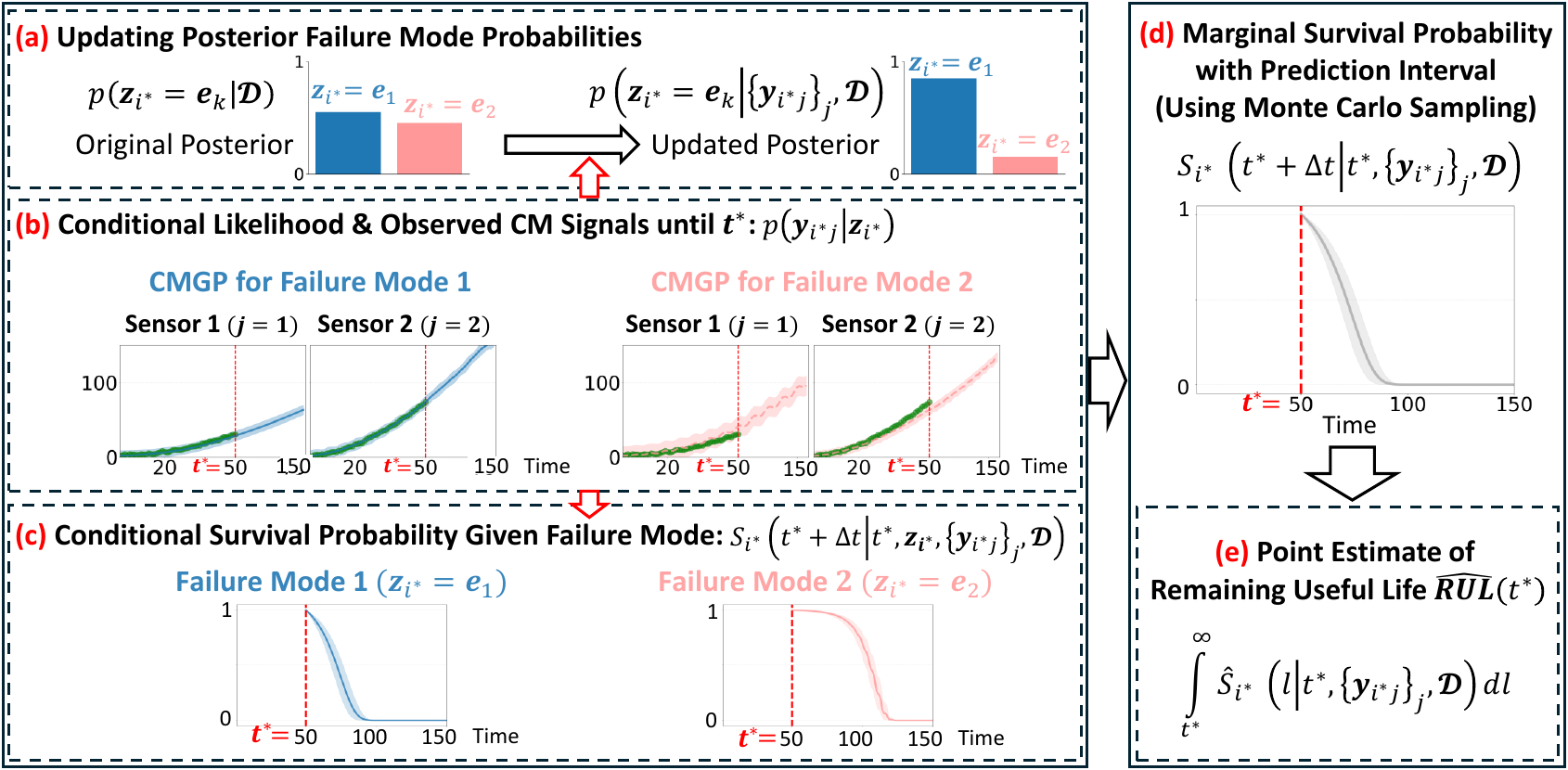}
	\caption{An overview of the prediction procedure in the proposed framework.}\label{fig:summary}
\end{figure}

The primary goal of the proposed model is to predict the failure modes and survival probabilities of a new unit $i^*$. When the unit is brand new, those probabilities are obtained only based on the historical data, and the probabilities are updated as new CM signals $\mbr{\y_{i^*j}}_j$ are obtained over time. An overview of the prediction procedure is illustrated in Figure~\ref{fig:summary}.

The posterior predictive survival probability for unit $i^*$ given CM signals $\mbr{\y_{i^*j}}_j$, \(S_{i^*}(t^* + \Delta t\mid t^*, \mbr{\y_{i^*j}}_j, \mathcal{D})\), is obtained by marginalizing over failure mode $\z_{i^*}$ (Figure~\ref{fig:summary}\hyperref[fig:summary]{(d)}):
\begin{align}
\label{eq:Marginal_survival_probability}
S_{i^*}(t^* + \Delta t&\mid t^*, \mbr{\y_{i^*j}}_j, \mathcal{D}) =\nonumber
\\
&\sum_{k=1}^{K} S_{i^*}(t^* + \Delta t \mid t^*, \z_{i^*} = \e_k, \mbr{\y_{i^*j}}_j, \mathcal{D}) p(\z_{i^*} = \e_k \mid \mbr{\y_{i^*j}}_j,\mathcal{D}).
\end{align}
where \( S_{i^*}(t^* + \Delta t \mid t^*, \z_{i^*} = \e_k, \mbr{\y_{i^*j}}_j, \mathcal{D}) \) is the conditional (posterior predictive) survival probability conditioned on each mode, and  \( p(\z_{i^*} \mid \mbr{\y_{i^*j}}_j, \mathcal{D}) \) is the posterior distribution of experiencing failure mode $\z_{i^*}$ given the observed CM signals up to time \(t^*\).

When no CM observations are available, the posterior distribution on failure modes $\z_{i^*}$ is obtained based solely on the historical data, as no information on the unit $i^*$'s failure modes is available; that is $p(\z_{i^*} \mid \mathcal{D})$.
The posterior distribution $p(\boldsymbol{\Pi}\mid \mathcal{D})$ is approximated by the variational distribution \(  
q(\boldsymbol{\Pi}) =  \text{Dir}(\bm{\tilde\alpha}^*)\), and thus, the posterior predictive distribution of $z_{i^*}$
is
\begin{align*}
p(\mathbf{z}_{i^*} = \mathbf{e}_k \mid \mathcal{D}) 
&\approx \int p(\mathbf{z}_{i^*} = \mathbf{e}_k  \mid \boldsymbol{\Pi}) \, q(\boldsymbol{\Pi} ) \, d\boldsymbol{\Pi} = \int \Pi_k \, q(\boldsymbol{\Pi}) \, d\boldsymbol{\Pi} = \mathbb{E}_{q}[\Pi_k] = \frac{\tilde{\alpha}^*_k}{\sum_{s=1}^{K} \tilde{\alpha}^*_s}.
\end{align*}
where \(\tilde{\alpha}^*_k\) denotes the optimized value of the variational parameter for \(q(\bm{\Pi})\).

When CM observations $\mbr{\y_{i^*j}}_j$ are obtained, the 
The posterior predictive distribution $p\sbr{\z_{i^*}=\e_k \mid \mbr{\y_{i^*j}}_j,\mathcal{D}}$ is updated by the Bayes' theorem (Figure~\ref{fig:summary}\hyperref[fig:summary]{(a)}):
\begin{equation}
\begin{aligned}
\frac{p\sbr{\z_{i^*} = \e_k, \mbr{\y_{i^*j}}_j\given\mathcal{D}}}{p\sbr{\mbr{\y_{i^*j}}_j\given\mathcal{D}}}
=
\frac{\prod_{j=1}^J p(\mathbf{y}_{i^* j} \mid \mathbf{z}_{i^*} = \mathbf{e}_k) \cdot p(\mathbf{z}_{i^*} = \mathbf{e}_k \mid \mathcal{D})}
{\sum_{r=1}^K \prod_{j=1}^J p(\mathbf{y}_{i^* j} \mid \mathbf{z}_{i^*} = \mathbf{e}_r) \cdot p(\mathbf{z}_{i^*} = \mathbf{e}_r \mid \mathcal{D})}.
\label{eq:class_prob}
\end{aligned}
\end{equation}
In Equation \eqref{eq:class_prob}, the posterior predictive distribution is updated based on the likelihood evaluation of the observed CM signals $\mbr{\y_{i^*j}}_j$ for the new unit \( i^* \), conditioned on each failure mode (i.e., \(p(\mathbf{y}_{i^* j} \mid \mathbf{z}_{i^*})\)). The probability of failure mode $k$ is updated higher when the CM observations $\mbr{\y_{i^*j}}_j$ are well aligned with the conditional CMGP prediction on given $k$ (i.e., higher likelihood \( p(\mathbf{y}_{i^* j} \mid \mathbf{z}_{i^*} = \mathbf{e}_k) \) from all sensors $j$), and lower otherwise (Figure~\ref{fig:summary}\hyperref[fig:summary]{(b)}).

Conditional posterior predictive survival function on failure mode \( S_{i^*}(t^* + \Delta t \mid t^*, \z_{i^*} = \e_k, \mbr{\y_{i^*j}}_j, \mathcal{D}) \) characterizes varying failure probabilities across different failure modes (Figure~\ref{fig:summary}\hyperref[fig:summary]{(c)}), by marginalizing out the posterior distributions of latent variables $b_k$, $\rho_k$, $f_{i^*j}$. 
\begin{equation}
\begin{aligned}
\widehat{S}_{i}(t^* + \Delta t \mid t^*, \z_{i^*}=\e_k, &\mbr{\y_{i^*j}}_j)=
\\
 &\iiint \exp \bigg\{ - \int_{t^*}^{t^* + \Delta t} \exp[b_k + \rho_k l] \exp \left[ \widehat{\boldsymbol{\gamma}}_k{}^\top \mathbf{x}_{i^*} + \sum_{j=1}^{J} \widehat{\beta}_{kj} f_{i^*j}(l) \right] dl \bigg\}
 \\
 &\times  {q}(b_{k})\, 
{q}(\rho_{k})\,
p({f}_{i^*j} \mid \z_{i^*} = \e_k, \mbr{\y_{i^*j}}_j)\,
d{f}_{i^*j}\, db_{k}\, d\rho_{k}
\label{eq:sk}
\end{aligned}
\end{equation}
where numerical integrations are performed by using Monte Carlo sampling, as Equation~\eqref{eq:sk} is analytically intractable;
\(f_{i^*j}\), \(b_k\), and \(\rho_k\) are sampled from the corresponding posterior distributions $p({f}_{i^*j} \mid \z_{i^*} = \e_k, \mbr{\y_{i^*j}}_j)$, ${q}(b_{k};\tilde{\mu}_{b_k}^*,\tilde{\sigma}^*_{b_k}{}^2)$, and ${q}(\rho_{k};\tilde{\alpha}^*_{\rho_k},\tilde{\beta}^*_{\rho_k})$, respectively (The starred parameters denote optimized values for the variational distributions.) 
The marginalization in \eqref{eq:sk} fully characterizes the uncertainties on CM signals ($f_{i^*j}|y_{i^*j},\dataset$) and baseline hazard parameters ($b_k,\rho_k|y_{i^*j},\dataset$), which is crucial to ensure precise uncertainty quantification and unbiased prediction. 
On the other hand, many existing prognostic models rely on plugging in the point prediction of the CM signal or do not consider priors for hazard parameters for survival prediction for simplicity \citep{yue2021joint, yu2004joint, zhou2014remaining,mauff2020joint, wen2023neural}.

Lastly, the marginalized posterior predictive survival function $S_{i^*}(t^* + \Delta t\mid t^*, \mbr{\y_{i^*j}}_j, \mathcal{D})$ in \eqref{eq:Marginal_survival_probability} is also then calculated by Monte Carlo sampling.
Detailed sampling schemes are included in Section S5.1.3 of the supplementary material.

The proposed method produces the sampled posterior predictive distributions of the survival function, from which point estimates $\widehat{S}_{i^*}$ and prediction intervals of survival functions can be easily obtained. Point estimate of RUL can also be easily obtained by integrating \(\widehat{S}_{i^*}\) from the decision time \( t^* \) onward (Figure~\ref{fig:summary}\hyperref[fig:summary]{.(e)})
\begin{equation}
    \widehat{\text{RUL}}_{i^*}(t^*) = \mathbb{E}(T_{i^*}-t^*\mid t^*) = \int_{t^*}^{\infty} \widehat{S}_{i^*}(l\mid t^*, \mbr{\y_{i^*j}}_j, \mathcal{D} ) \, dl,
    \label{eq:mrl}
\end{equation}
which represents the expected remaining time to failure, given that the unit has survived up to time \( t^* \).

The whole prediction mechanism is clearly illustrated in Figure~\ref{fig:summary}(a)-(e). 
Without observations, the posterior distribution of $\z_i$ accounts for the historical failure mode probabilities (left bar graphs in Figure~\ref{fig:summary}\hyperref[fig:summary]{(a)}).
When CM signals observed (green points in Figure~\ref{fig:summary}\hyperref[fig:summary]{(b)})
from a new unit \(i^*\), they are evaluated based on the conditional CM likelihood (CMGP posterior distributions on $\f$ given $\z$).
In this example, the observations are better aligned with the CMGP corresponding to failure mode 1, resulting in a higher conditional CM likelihood for the mode 1, and uplifting the posterior distribution on $\z_i$ with a higher probability of mode 1 (Right bar graphs in Figure~\ref{fig:summary}\hyperref[fig:summary]{(a)}). New CM signals also update the conditional survival probability given failure modes (Figure~\ref{fig:summary}\hyperref[fig:summary]{(c)}). By integrating both distributions on failure time and mode, the (marginalized) posterior predictive survival probability is updated (Figure~\ref{fig:summary}\hyperref[fig:summary]{(d})), and the point estimate RUL can be easily obtained from it (Figure~\ref{fig:summary}\hyperref[fig:summary]{(e)}).

\section{Validation}
\label{sec:Validation}

Our proposed model is validated by extensive numerical and case studies, using multiple metrics to evaluate various aspects.

\subsection{Methods Used for Validation}

The performance of our proposed model \ours is compared with four benchmark models:

\begin{itemize}

    \item \deepbrch: A fully neural network-based model that uses LSTM networks in combination with classification and regression networks ~\citep{9845469}. \deepbrch makes point prediction of both failure mode and RUL and does not quantify the uncertainty.

    \item \nnjoint: A hybrid model where a linear mixed-effects model predicts CM signals, which are input to a neural network predicting the hazard function without taking failure mode into account \citep{wen2023neural}. \nnjoint predicts survival probability with partial uncertainty quantification, only characterizing the uncertainties in the linear model, but not the uncertainty of the survival model.
    
    This model is not designed for multiple failure modes; therefore, for fair comparison, we extend the neural network by adding classification nodes for failure mode prediction (See Section~S4.1 of the supplementary material for more details). In the numerical study, we use the exact true basis functions for the linear mixed-effects component, as they are known. For the case study, we adopt a quadratic form, as proposed in \cite{wen2023neural}.

    \item \deepsurv: A Cox PH model where its hazard function is approximated by the neural network \citep{katzman2018deepsurv}.
    \deepsurv does not account for time-variant covariates (CM signals) and multiple failure modes; we incorporate the most recent observation of CM signals, also known as the Last-Observation-Carried-Forward (LOCF) method~\citep{cox1984analysis,arisido2019joint}, commonly used for handling time-dependent covariates in CoxPH and DeepSurv~\citep{wen2023neural,huh2024integrated}. Additionally, the neural network is extended for a failure mode prediction, similarly to \nnjoint.

    \item \coxph: Classical survival model; \coxph does not consider multiple failure modes, time-dependent covariates (CM signals), and uncertainty quantification \citep{cox1972regression}. 
    LOCF is used to incorporate the CM signals. \coxph predicts the survival function without taking into account different failure modes.

\end{itemize}

\subsection{Evaluation Metrics}
\label{ssec:metrics}

We evaluate three aspects of the aforementioned methods, where applicable.

\begin{itemize}

\item \textbf{Failure Mode Prediction}: {Prediction performance of failure mode is assessed by prediction errors: \(
1-p(\mathbf{z}_{i^*} = \mathbf{e}_k \mid \mbr{\y_{i^*j}}_j,\mathcal{D}) 
\) for unit \(i^*\) with true failure mode \(k\).
}

\item \textbf{RUL Prediction}: The performance of RUL point prediction is assessed by the absolute errors: \(
\left| \widehat{\text{RUL}}_{i^*} - \text{RUL}_{i^*} \right|
\) for unit \(i^*\), where $\widehat{\text{RUL}}_{i^*}$ denotes the predicted expected RUL. For $\text{RUL}_{i^*}$, the true expected RUL is used in the numerical studies, and the observed failure times are used in the case study where the true expected values are not available.

\item \textbf{Uncertainty Quantification}: Quality of the predicted credible intervals for the survival functions is evaluated in the numerical studies by the coverage proportion that the predicted credible interval includes the true survival probability, $S^{True}_{i^*}$.
\begin{equation}
\begin{aligned}
\label{eq:MCR}
\text{Coverage Proportion} = \frac{1}{I^*} \sum_{i^*=1}^{I^*} \left( \frac{1}{N} \sum_{r=1}^N \mathbb{I}\{ S^{True}_{i^*}(t^* + \,\Delta r \mid t^*) \in [\hat{L}_{ir}, \hat{U}_{ir}] \} \right),
\end{aligned}
\end{equation}
where \(I^*\) is the total number of test units, \(N\) is the total number of predicted time points, \(\Delta\) is their interval, and \(\mathbb{I}\{\cdot\}\) is an indicator function that equals one if the the prediction interval \([\hat{L}_{ir}, \hat{U}_{ir}]\) includes the true survival probability and zero otherwise.

\end{itemize}

\subsection{Numerical Study}
We have conducted an extensive numerical study with the simulated dataset to demonstrate the advantages of the proposed model. 

\subsubsection{Data Generation}
Failure times and CM signals from two sensors ($J=2$) are simulated under two distinct failure modes ($K=2$). 
The signals are generated according to the model \( y_{kij}(t) = \mathbf{Z}_{kj}(t)^{\top}\,\mathbf{B}_{kij} + \epsilon_{kj}(t) \), where the random coefficient column vectors \( \mathbf{B}_{kij} \) are drawn from a multivariate normal distribution, and the noise term follows \( \epsilon_{kj}(t) \sim \mathcal{N} \bigl(0, \sigma_{kj}^2 \bigr) \), similar to \cite{zhou2014remaining}, \cite{yue2021joint}, \cite{wen2023neural}, and \cite{huh2024integrated}. The basis functions \( \mathbf{Z}_{kj}(t) \) are designed for nonlinear degradation patterns over time, different across sensors and failure modes.
The failure times are then generated by using the hazard function, given as \(
h_{ki}(t) = h_{0k}(t)\,\exp\bigl(\sum_{j=1}^{J} \beta_{kj}\,\mathbf{Z}_{kj}(t)^\top \mathbf{B}_{kij} \bigr),\) for each unit $i$
where \( \beta_{kj} \) denotes the regression coefficient specific to failure mode \( k \) and sensor \( j \).
Detailed configurations of data generation in the numerical studies are provided in Section~4.2 of the supplementary material. For training units, we draw 50 samples of failure times \(T_i\) for each of two failure modes ($I=100$) from the failure time density \(f_i(t) = h_i(t) S_i(t)\) using rejection sampling. 
Similarly, for testing units, we generate 10 samples for each failure mode.

\begin{figure}[ht!]
\centering
\includegraphics[width=0.8\textwidth]{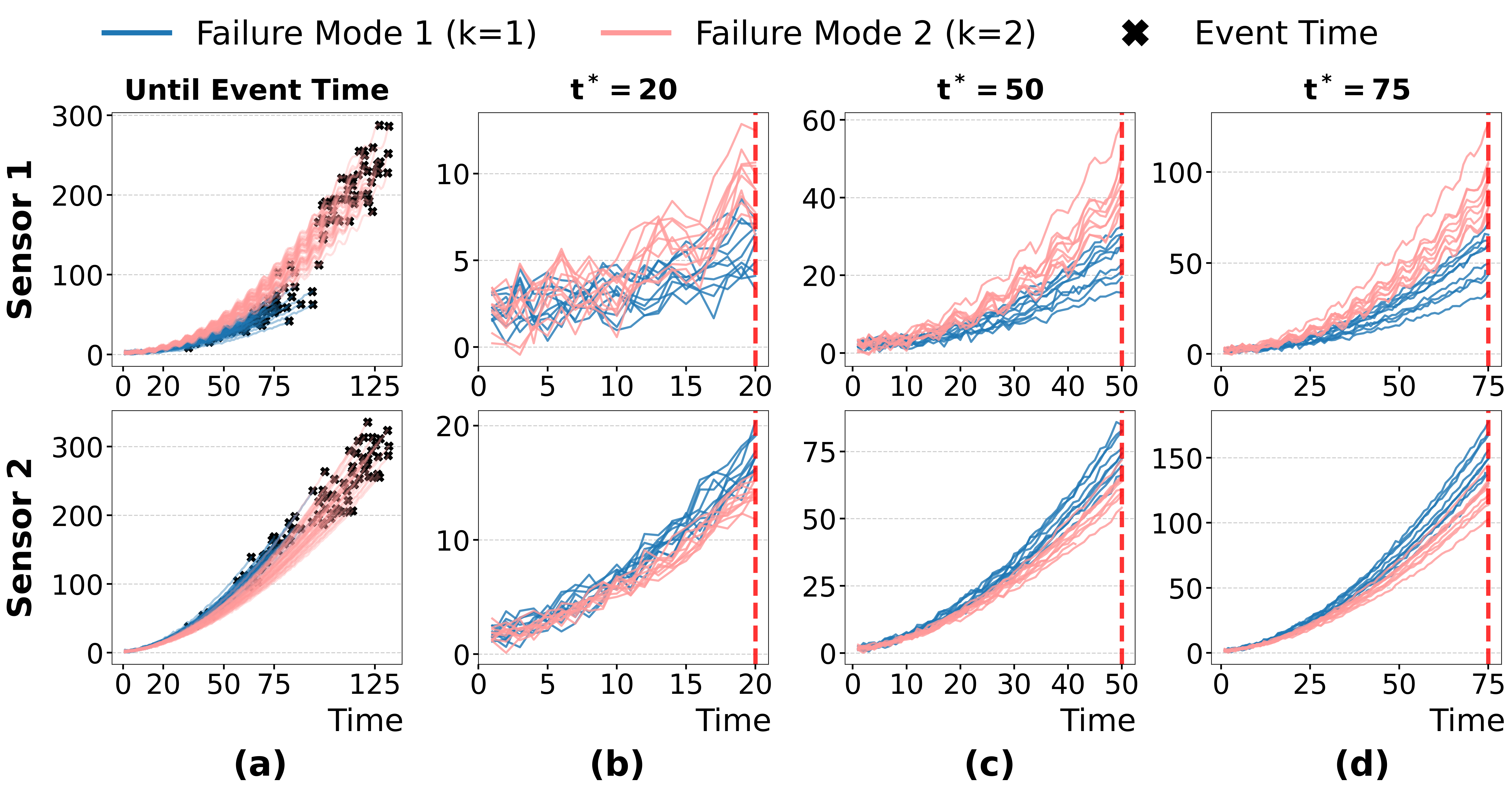}
\caption{Generated condition monitoring signals for the numerical study.
 Historical CM observations until failure are shown in (a). Test units CM observations up to different $t^*$ are shown in (b) $t^*=20$, (c) $t^*=50$, and (d) $t^*=75$. 
}
\label{fig:NS_Generated_data}
\end{figure}

In prediction, sensor observations are observed up to three different times \( t^* (=20, 50, 75) \) and the prediction performance is evaluated by the given metrics shown in Section \ref{ssec:metrics}.
Figure \ref{fig:NS_Generated_data} illustrates the generated CM observations.
Figures~\ref{fig:NS_Generated_data}\hyperref[fig:NS_Generated_data]{(a)} depicts the CM observations and the event time in the training data, and Figure~\ref{fig:NS_Generated_data}\hyperref[fig:NS_Generated_data]{(b)},  
~\ref{fig:NS_Generated_data}\hyperref[fig:NS_Generated_data]{(c)}, and  
~\ref{fig:NS_Generated_data}\hyperref[fig:NS_Generated_data]{(d)}  show the test CM observations up to \( t^* = 20, 50, 75 \). 

At earlier stages (e.g., $t^*=20$), the CM signals from different failure modes exhibit significantly higher overlaps with unclear distinction. Therefore, it is very challenging to identify the underlying failure mode based on these limited CM observations.

\subsubsection{Failure Mode and RUL Prediction Performance}
\label{ns_ft}

The performance of the proposed method is evaluated in comparison to benchmarks with respect to failure mode prediction, RUL prediction, and uncertainty quantification.
Our proposed model outperforms benchmark methods in all three key aspects. 
Furthermore, \ours exhibits a significant advantage in early-stage prediction where only limited CM signals are observed. 

\begin{figure}[ht!]
  \centering
  \begin{minipage}[t]{0.48\textwidth}
    \centering
    \includegraphics[width=\textwidth]{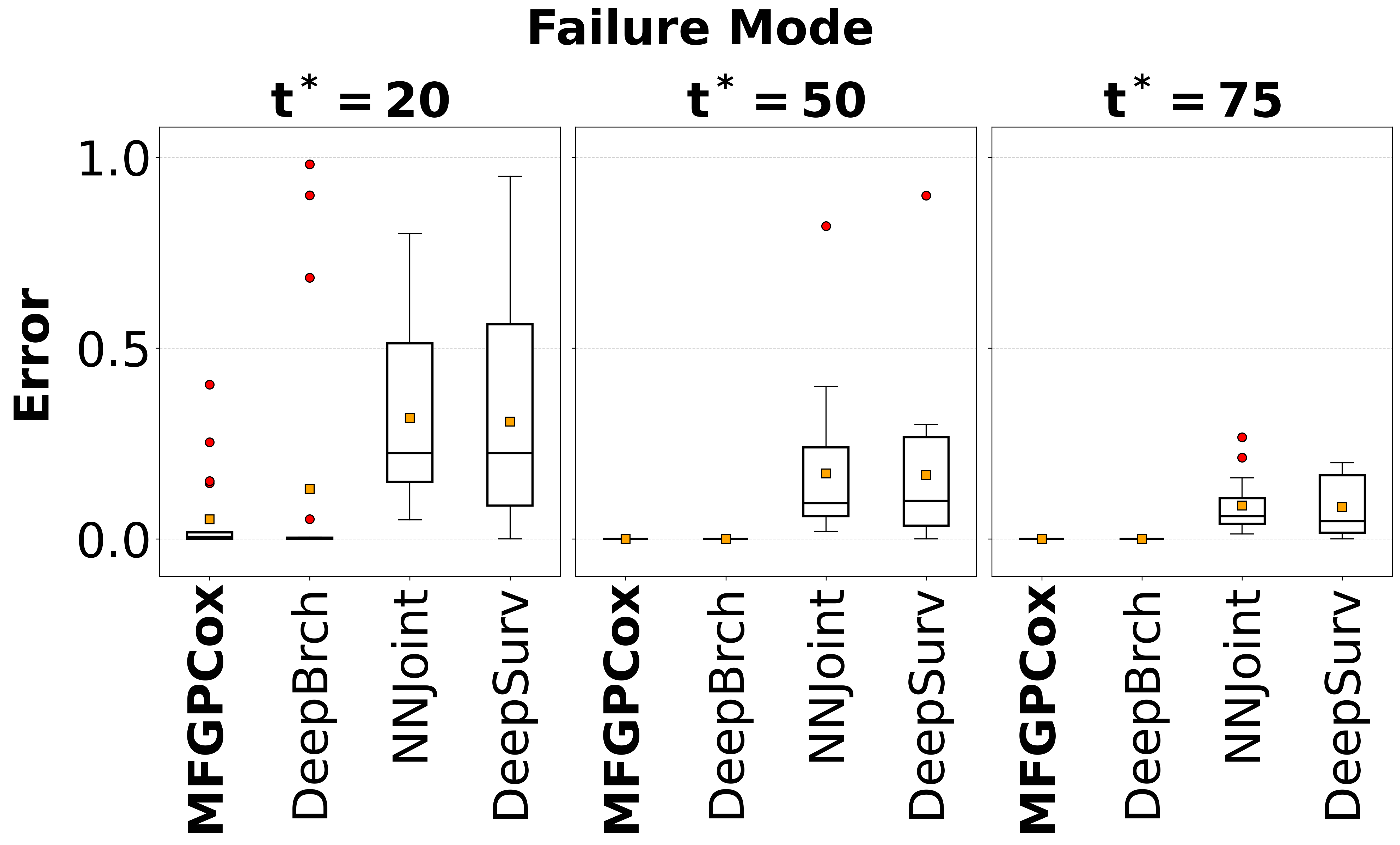}
    \caption{Errors of test units failure mode predictions in the numerical study.}
    \label{fig:F1_score}
  \end{minipage}%
  \hfill
  \begin{minipage}[t]{0.48\textwidth}
    \centering
    \includegraphics[width=\textwidth]{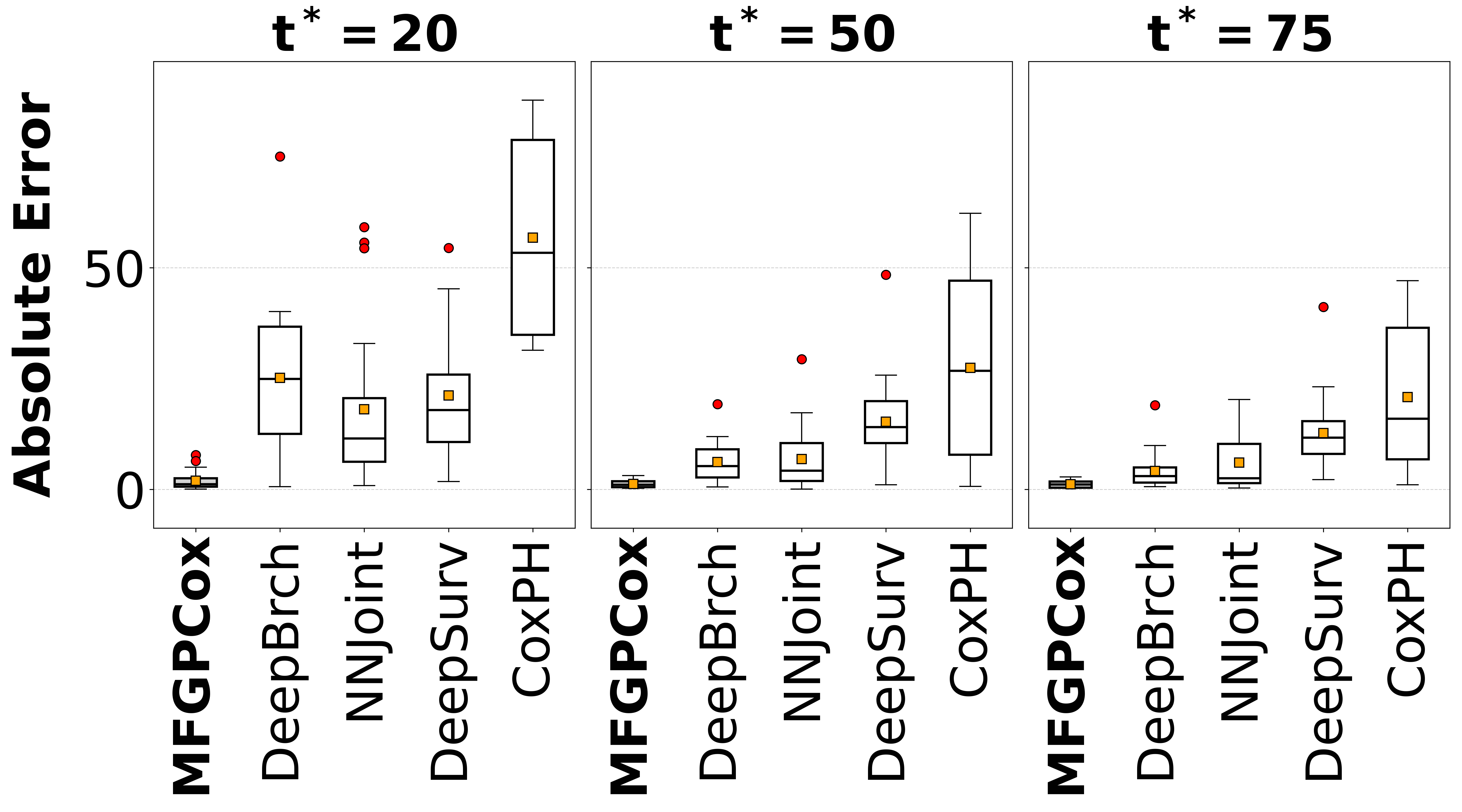}
    \caption{Absolute errors of test units RUL predictions in the numerical study.}
    \label{fig:NS_AE_RUL}
  \end{minipage}
\end{figure}

Figures~\ref{fig:F1_score} and \ref{fig:NS_AE_RUL} depict the errors for failure mode and RUL prediction, respectively. 
\coxph is not included for failure mode prediction evaluation, as it does not account for failure modes, but is evaluated in its RUL prediction. Due to the association between failure mode and time, RUL prediction is more accurate when failure mode prediction is more accurate with higher confidence.
The figures confirm the benefits of the proposed model \ours, demonstrating its superior performance in both failure mode and RUL prediction with the lowest mean errors across all \(t^*\). When the predicting time $t^*$ increases, the predictive errors tend to decrease in all the methods.

At \(t^* = 20\), due to the limited CM observations with higher overlaps observed in Figure~\ref{fig:NS_Generated_data}\hyperref[fig:NS_Generated_data]{(b)} in the test units, failure modes prediction naturally involves greater uncertainty. Both \ours and \deepbrch show overall low errors in failure mode prediction; however \deepbrch is overconfident when it is wrong. On the other hand, \nnjoint and \deepsurv exhibit significantly higher errors at $t^* = 20$. 
\ours and \deepbrch become perfectly correct in failure mode prediction since $t^*=50$, while \nnjoint and \deepsurv become increasingly improved, but some errors remain even at $t^*=75$.  \ours has the lowest errors in RUL prediction, which even gets improved over time. 
\deepbrch show higher errors than \nnjoint and \deepsurv at $t^* = 20$, but improve over time and finally produce lower errors than \nnjoint and \deepsurv at $t^*=75$. \nnjoint and \deepsurv also improve over time, but their improvement were slower than \deepbrch's. CoxPH generally shows the lowest performance in RUL prediction.

The significant outperformance of the proposed \ours model with robustness results from advantageous features of the model: 1) Each data types (i.e., failure time, CM signals, and failure modes) is accounted for by precise probabilistic models (i.e., Cox PH model, CMGP, and categorical distributions), capturing individual characteristics as well as the similarities across units in a hierarchical Bayesian framework, where 2) their parameters are estimated simultaneously in a unified model and 3) multiple sensor data are effectively learned (See Section S4.3 in the supplementary material).
The holistic view alongside an individualized perspective enables more precise and accurate failure mode and RUL predictions.
Bayesian nature of the model facilitates highly efficient and effective prediction by considering the whole possibilities with the weights of quantified uncertainties given the observations, even when they are limited at early stages.

\deepbrch also provides the holistic view with a neural network framework, specifically, LSTM networks. With the comprehensiveness of the model, it outperforms the other two neural network-based models. \deepbrch generally shows lower performance at early prediction, whereas its prediction at a later stage is significantly better than the other two neural network-based models. However, neural network-based models tend to be overconfident in their predictions. Additionally, it does not directly account for any types of uncertainty, only providing a point prediction of RUL, unable to predict the probability distribution of failure time or survival function. 
\nnjoint and \deepsurv, on the other hand, can predict the survival functions, considering some uncertainties. However, the CM signals are not accounted for within the survival prediction model. \nnjoint uses an independent parametric model that oversimplifies the CM signals, and \deepsurv uses the latest observation (LOCF) of CM signals. While both models underperform \deepbrch, they provide richer outcomes by predicting probabilities of failure (survival functions) and \nnjoint characterizes uncertainty of CM signals through a linear model. 
In contrast, \ours addresses the limitations of these models by (1) comprehensive model, inference, and prediction, (2) uncertainty quantification, (3) nonparametric modeling, and (4)  multiple failure modes.

\subsubsection{Uncertainty Quantification}
\label{sssec:UQ-NS}

Another advantage of our proposed model is that it fully characterizes the uncertainties of data and models by the fully Bayesian framework. The quality of uncertainty quantification is assessed by the proportion of the 95\%-credible interval's inclusion of the true survival probabilities, which ideally needs to be close to 0.95. 

\begin{wrapfigure}{r}{0.48\textwidth} 
    \vspace{-2em}
    \centering
    \includegraphics[width=0.46\textwidth]{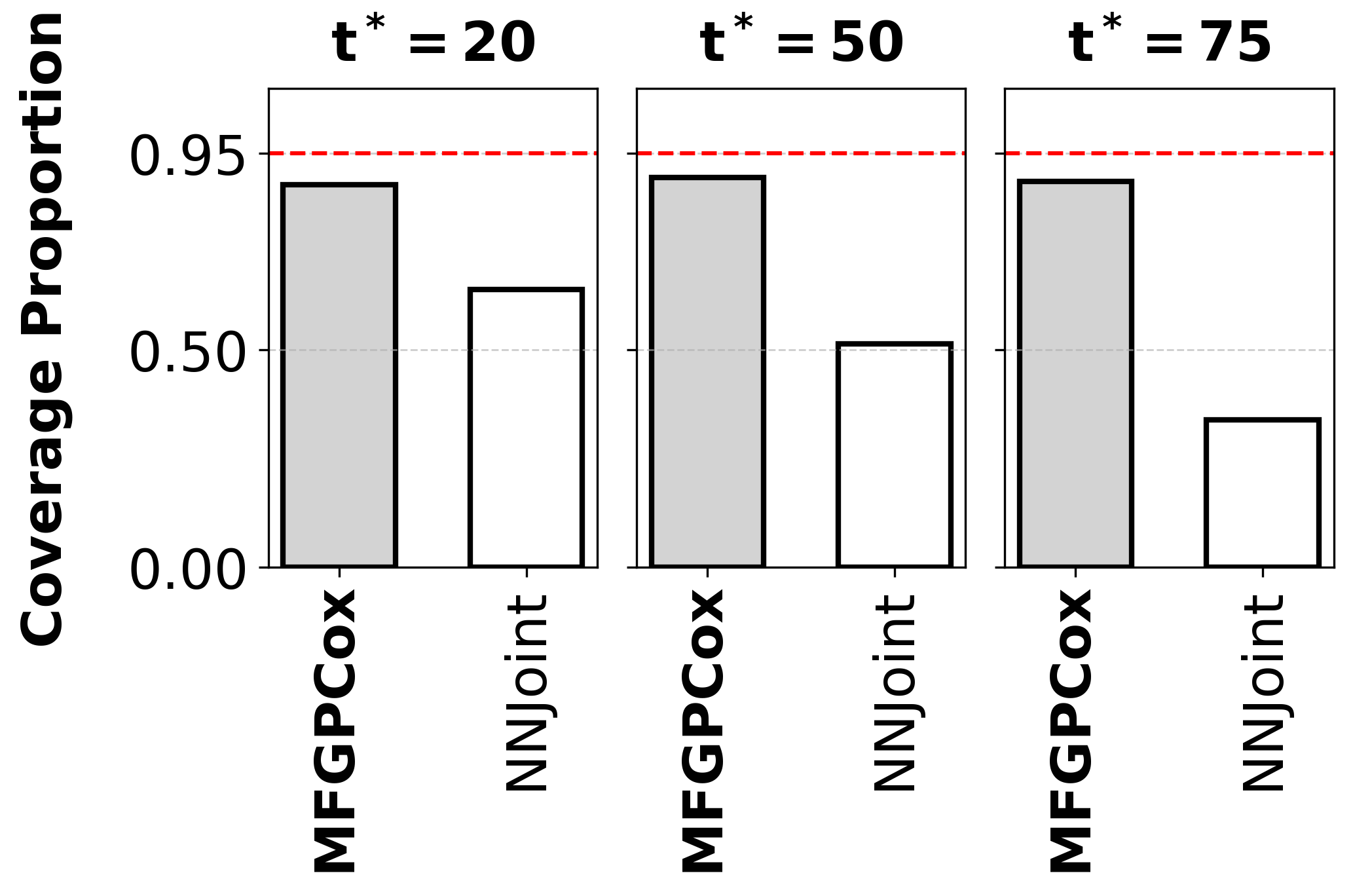}
 \caption{Comparison of coverage proportion between MFGPCox and NNJoint across all test units in the numerical study.}
\label{fig:mean_coverage_ratio}
\end{wrapfigure}

Figure~\ref{fig:mean_coverage_ratio} exhibits that the inclusion proportions of \ours are close to 0.95 in all three cases of $t^*=20, 50, 75$. The only benchmark model that quantifies the model uncertainty is \nnjoint, which is through the linear mixed effects model. However, the inclusion proportion of \nnjoint is much lower than 0.95. This is because (1) the linear model is independent of the survival model, (2) its parametric form is too rigid, and (3) the survival model is not modeled as Bayesian, failing to quantify the model uncertainties of the survival model.

\begin{figure}[ht!]
  \centering
  \begin{minipage}[t]{0.55\textwidth}
    \vspace{0pt}  
    \includegraphics[width=\textwidth]{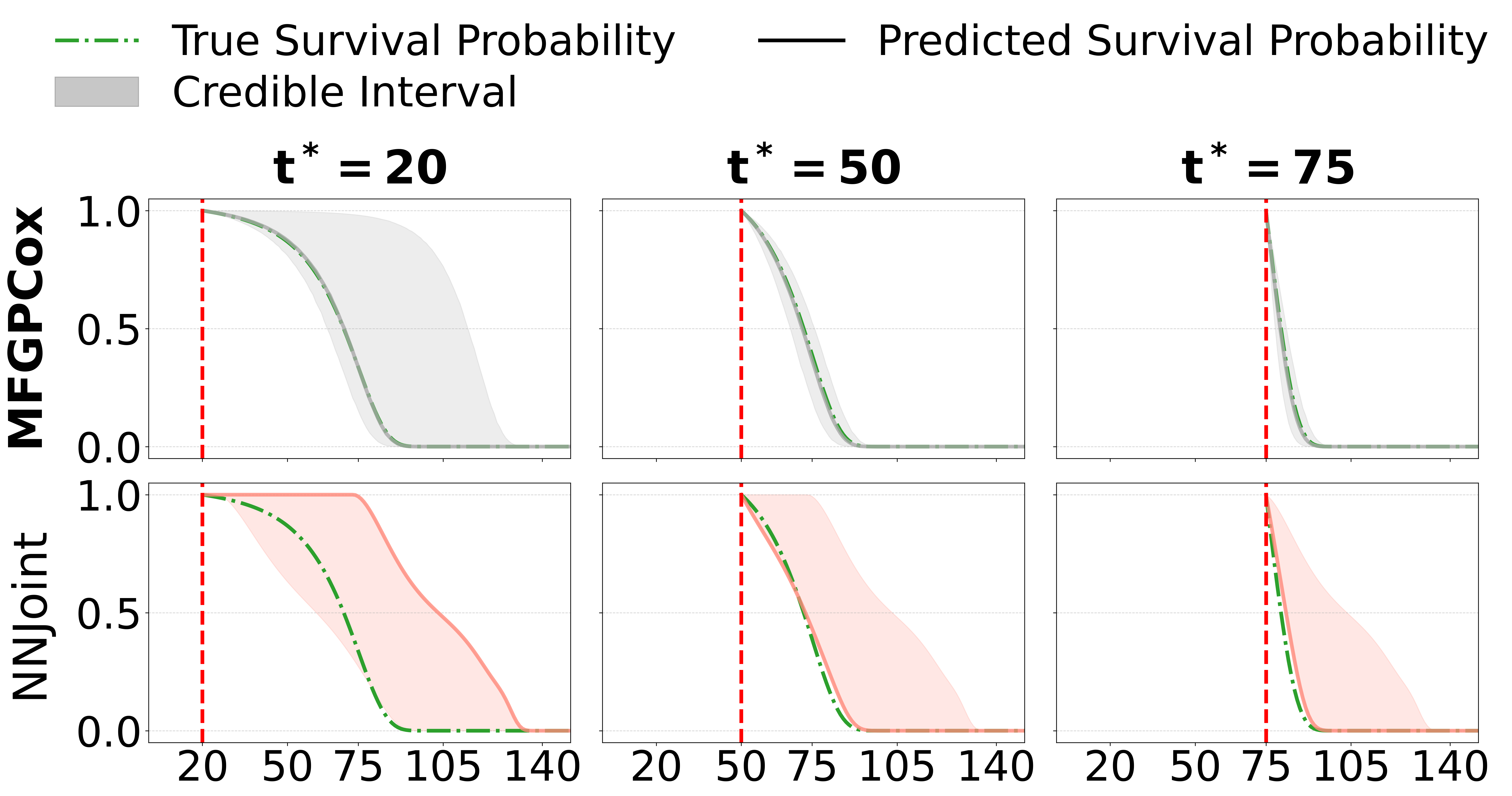}
    \caption{Predicted survival probabilities with credible interval for the selected test unit in the numerical study.}
    \label{fig:Survival_Probabilities_Grid_55}
  \end{minipage}%
  \hfill
  \begin{minipage}[t]{0.42\textwidth}
    \vspace{0pt}  
    \captionsetup{type=table}
    \caption{Predicted failure mode probabilities for the selected test unit at different time points (\( k = 1 \) denotes failure mode 1 and \( k = 2 \) denotes failure mode 2).}
   \renewcommand{\arraystretch}{0.35}
    \setlength{\tabcolsep}{3.5pt}  
    \begin{tabular}{c|cc|cc}
      \toprule
      \multirow{2}{*}{Time \( t^* \)} & \multicolumn{2}{c|}{MFGPCox} & \multicolumn{2}{c}{NNJoint} \\
       & \( k = 1 \) & \( k = 2 \) & \( k = 1 \) & \( k = 2 \) \\
      \midrule
      20  & 0.746 & 0.254 & 0.45 & 0.55 \\
      50  & 1.000 & 0.000 & 0.76 & 0.24 \\
      75  & 1.000 & 0.000 & 0.84 & 0.16 \\
      \bottomrule
    \end{tabular}
    \label{tab:failure_mode_probs}
  \end{minipage}
\end{figure}

Figure~\ref{fig:Survival_Probabilities_Grid_55} presents an illustrative example of a predicted survival function of one of the test units, whose true failure mode is \( k = 1 \).
It shows the predicted survival probabilities and credible intervals from both \ours and \nnjoint. At \( t^* = 20 \), both models show a wide credible interval due to uncertainty in failure mode prediction, which are shown in Table \ref{tab:failure_mode_probs}.  At \( t^* = 50 \) and \( t^* = 75 \), the credible interval of \ours has reduced significantly as the failure mode prediction becomes more precise, and the credible interval includes the true survival function (green curves). In contrast, \nnjoint's credible intervals remain high even at $t^*=75$ and fail to include the true survival function. 
This is because \nnjoint is (1) highly confident in predicting the CM signals from the linear model, leading to highly confident survival prediction, but less precise, and (2) the failure mode is predicted with lower confidence.

\subsection{Case Study}
\label{subsec:CS}

We also perform a case study to validate the performance in practical scenarios.

\subsubsection{Data Description}
\begin{wrapfigure}{r}{0.45\textwidth} 
    \centering
     \includegraphics[width=0.38\textwidth]{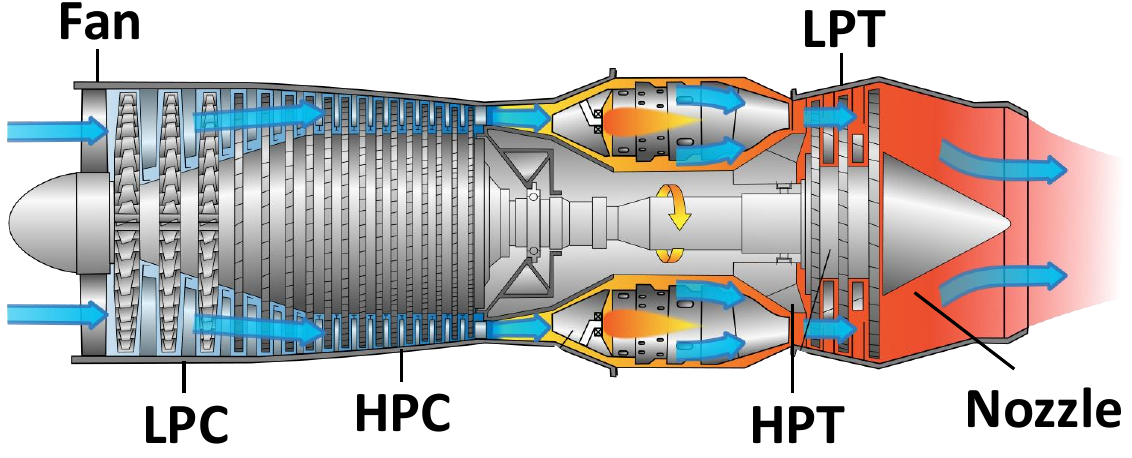}
    \caption{Simplified diagram of the simulated engine.}
    \label{fig:engine_schematic}
\end{wrapfigure}

The Commercial Modular Aero-Propulsion System Simulation (C-MAPSS) dataset, developed by NASA, is a widely used dataset for prognostics validation. It is generated by simulating turbofan engine operations (shown in Figure~\ref{fig:engine_schematic}) but does not include labels of the engines' failure modes \citep{frederick2007user, saxena2008turbofan}. 

\begin{figure}[ht!]
  \centering
  \begin{minipage}[t]{0.4\textwidth}
    \vspace{0pt}  
    \captionsetup{type=table}
    \caption{Description of the simulated condition monitoring signals in the case study dataset.}
    \renewcommand{\arraystretch}{0.6} 
    \setlength{\tabcolsep}{3pt}  
    \begin{scriptsize}
    \begin{tabular}{llc}
        \toprule
        Symbol & Description & Units \\
        \midrule
        NL & Low-pressure (LP) shaft speed & rpm \\
        NH & High-pressure (HP) shaft speed & rpm \\
        P13 & Fan outer outlet pressure & bar \\
        P26 & HP compressor inlet pressure & bar \\
        P3  & HP compressor outlet pressure & bar \\
        T3  & HP compressor outlet temperature & °C \\
        T6  & Exhaust gas temperature & °C \\
        T42 & HP turbine outlet temperature & °C\\
        \bottomrule
    \end{tabular}
    \end{scriptsize}
    \label{tab:sensor_data}
  \end{minipage}%
  \hfill
  \begin{minipage}[t]{0.55\textwidth}
    \vspace{0pt}  
    \includegraphics[width=\textwidth]{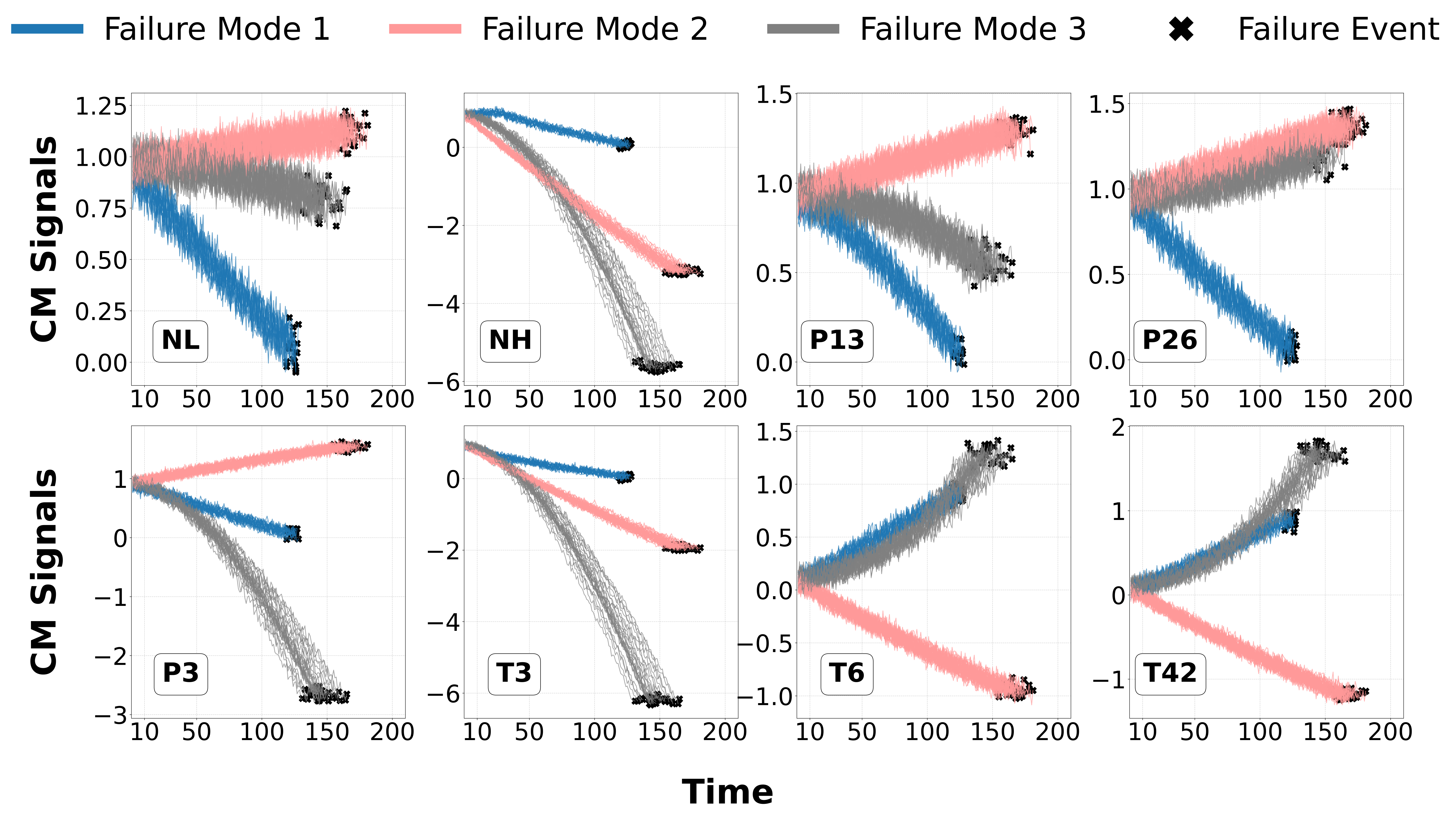}
    \caption{Selected condition monitoring signals from the case study dataset.}
    \label{fig:CS_8s}
  \end{minipage}
\end{figure}

Therefore, \cite{9845469} directly used the simulator TEACHES (Turbine Engine Advanced Calculation and Health Assessment Educational Software) to generate a dataset with labels of failure modes. This dataset includes CM signals from sixteen sensors attached to various components of engines and continuously monitors various factors \citep{9845469}. Some sensor signals show a clear distinction across failure modes even at an early stage. For robust validation, we have selected eight sensors that are more challenging in distinguishing between at least two of the three failure modes in an early stage \(t^* = 10\). Table~\ref{tab:sensor_data} describes the selected CM signals, which are depicted in Figure~\ref{fig:CS_8s}.

The dataset contains 150 units ($I=150$): 30, 70, and 50 from failure modes 1, 2, and 3, respectively. In training, we used 10 units from each failure mode, while the remaining 120 units were used as test units. The test data were truncated at \( t^* = 10, 25, \) and \( 75 \), at each of which prediction performance is assessed. 

\subsubsection{Failure Mode and RUL Prediction}
\label{sssec:Failure mode prediction-CS}

\begin{figure}[ht!]
  \centering
  \begin{minipage}[t]{0.48\textwidth}
    \centering
\includegraphics[width=\textwidth]{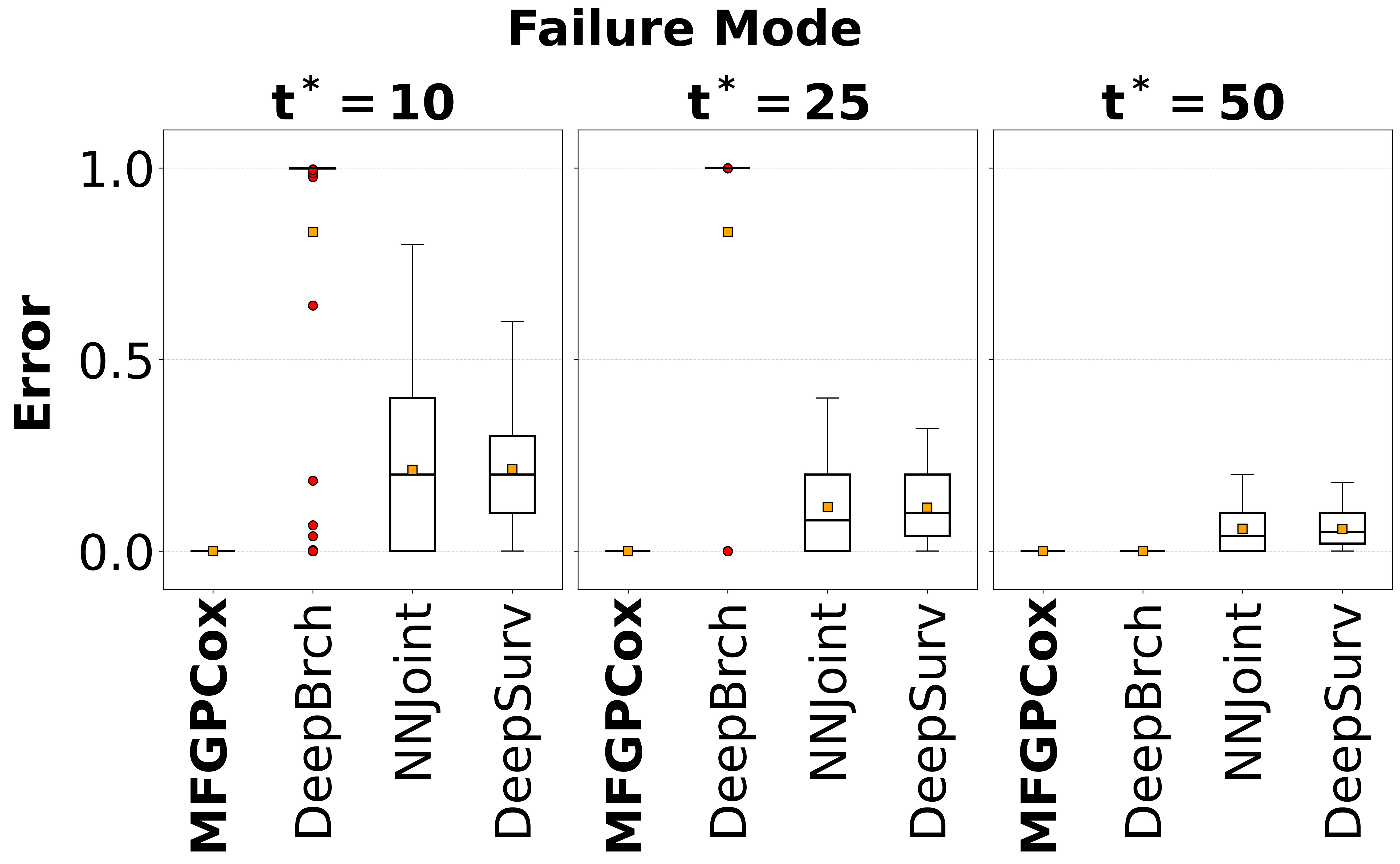}
  \caption{Errors of test units failure mode predictions in the case study.}
  \label{fig:CS_failure_mode_errors}
  \end{minipage}%
  \hfill
  \begin{minipage}[t]{0.48\textwidth}
    \centering
\includegraphics[width=\textwidth]{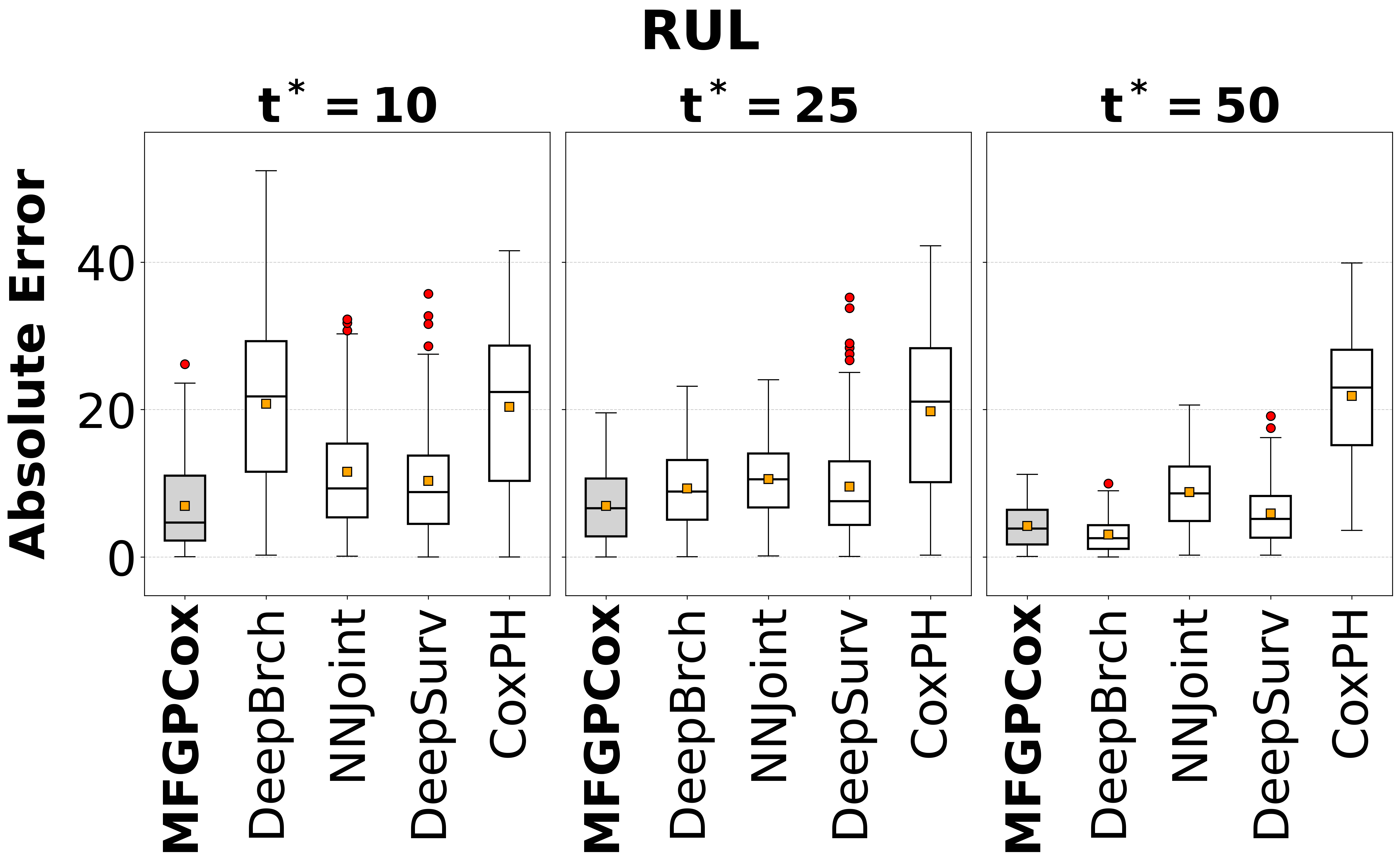}
  \caption{Absolute errors of test units RUL predictions in the case study. }
  \label{fig:CS_AE}
  \end{minipage}
\end{figure}

The results from the case study are consistent with the numerical study's, where our proposed \ours outperforms benchmark methods in both failure mode and RUL predictions. 
Figures~\ref{fig:CS_failure_mode_errors} and \ref{fig:CS_AE} depict the prediction errors of failure mode and RUL, respectively.
\ours excels especially at early-stage prediction ($t^*=10, 20$), demonstrating a clear advantage when observed data are limited.
All models, including \ours, improve in their performance when prediction time $t^*$ increases with more CM signals.

Figure~\ref{fig:CS_failure_mode_errors} highlights the superior performance of \ours, with perfect prediction across all \(t^*\). \deepbrch performs poorly at \(t^* = 10\) and \(25\), where it demonstrated significant over-confidence even when its prediction was inaccurate. However, the prediction was improved significantly at \(t^* = 50\). 
\nnjoint and \deepsurv show similar performance, whose prediction was better than \deepbrch at \(t^* = 10\) and \(25\) but fail in perfect prediction at $t^*=50$.
In figure \ref{fig:CS_AE}, \deepbrch's RUL prediction is inaccurate at \(t^* = 10\), whereas its performance improves significantly over time.  
\nnjoint and \deepsurv show similar performance, where \deepsurv slightly outperforms \nnjoint in the case study, especially at $t^*=50$, although \nnjoint slightly outperforms \deepsurv in the numerical study. This is because the true basis functions are used for the linear mixed effects model of \nnjoint in the numerical study. However, in the case study, quadratic basis functions are used as proposed in \cite{wen2023neural} because the underlying basis function is unknown. Such a mismatch leads to low performance. As \deepsurv and \coxph use the LOCF approach, their predictions improve as the unit approaches failure time. Similar to the numerical study, CoxPH shows the lowest performance among all models.

The uncertainty quantification is not assessed for the case study as the underlying survival function is not available; instead, an illustration of the predicted credible intervals for the case study is provided in Section~S4.4 of the supplementary material.

\section{Conclusion}
\label{sec:conc}

In this paper, we proposed a unified hierarchical Bayesian model for jointly predicting failure times and modes of industrial systems subject to multiple modes of failure based on real-time CM signals. Three data sources, failure times and modes and CM signals are represented by Cox PH, multinomial, and CMGP, which are seamlessly integrated in Bayesian framework, where CMGP and Cox PH effectively characterize individual variations while capturing shared characteristics. The posterior distributions of the entire latent variables of the model are obtained by MFVI for effective inference. Such comprehensive view leads to less biased predictions. The proposed model concurrently predicting both failure mode and time with corresponding quantified uncertainty, performed by Monte Carlo sampling to enhance statistical rigors.
In the extensive numerical and case studies, we highlighted the significant advantages of our proposed models over benchmarks in all failure mode prediction, failure time (RUL) prediction, and uncertainty quantification.

While the proposed model is highly effective, it also offers opportunities remaining for further investigation and future research. Despite the approximation, the CMGP still remains computational demanding with massive dataset size. Also, different data modes are often available, such as images or frequency-domain representations of time-series signals. Leveraging the flexibility of the modular structure of the proposed framework, one potential solution is to replace the CMGP component with a Bayesian neural network, which offers greater scalability and efficiency for handling large volumes of data while it can still providing uncertainty quantification. Bayesian neural networks are flexible to be extended for multimodal data integration.

\bigskip
\begin{center}
{\large\bf SUPPLEMENTARY MATERIALS}
\end{center}

\begin{description}
\item[Section S1: Derivation of the CMGP Kernels]: A detailed mathematical derivation of the CMGP kernels is provided.
\item[Section S2: Derivation of CMGP Posterior Predictive Distribution]: A detailed mathematical derivation of the CMGP posterior predictive distribution is provided.
\item[Section S3: Derivation of the Evidence Lower Bound (ELBO)]: Detailed derivations of the evidence lower bound for variational inference are provided.
\item[Section S4: Supplementary Materials for the Validation Section]: Data generation settings for the numerical studies, the advantages of the proposed method in leveraging multiple sensor signals, and uncertainty quantification in the case study are discussed.
\item[Section S5: Implementation Details]: Detailed implementation configurations are discussed.
\end{description}

\bibliographystyle{chicago}
\begingroup
    \setstretch{1}

\endgroup


\end{document}